\documentclass[12pt,preprint]{aastex}
\usepackage{bm}  
\usepackage{soul} 
\usepackage{amsmath}

\newcommand{\pivec}{\mbox{\boldmath $\pi$}}
\newcommand{\muvec}{\mbox{\boldmath $\mu$}}

\newcommand{\myemail}{jskowron@astrouw.edu.pl}
\newcommand{\cf}{cf.}

\newcommand{\bmu}{{\bm{\mu}}}

\newcommand{\e}{{\rm{E}}}
\newcommand{\mas}{\mathrm{mas}}
\newcommand{\masyr}{\mathrm{mas}\,\mathrm{yr}^{-1}}
\newcommand{\thetae}{\theta_\e}
\newcommand{\kpc}{\mathrm{kpc}}
\newcommand{\kms}{\mathrm{km}\,\mathrm{s}^{-1}}

\shorttitle{Neptune-class planet in Galactic bulge}
\shortauthors{Skowron et al.}

\begin{document}

\title{MOA 2011-BLG-028L\MakeLowercase{b}: a Neptune-mass Microlensing Planet in the Galactic Bulge$\,^\ast$}


\author{
J.~Skowron\altaffilmark{O1},           
A.~Udalski\altaffilmark{O1},        
R.~Poleski\altaffilmark{O1,O2},           
S.~Koz{\l}owski\altaffilmark{O1}, 
M.~K.~Szyma{\'n}ski\altaffilmark{O1},  
{\L}.~Wyrzykowski\altaffilmark{O1},    
K.~Ulaczyk\altaffilmark{O4}            
P.~Pietrukowicz\altaffilmark{O1},      
G.~Pietrzy{\'n}ski\altaffilmark{O1},   
I.~Soszy{\'n}ski\altaffilmark{O1}\\    
(The OGLE Collaboration$^{O}$)\\
        and\\
F.~Abe\altaffilmark{M5}, 
D.~P.~Bennett\altaffilmark{M2}, 
A.~Bhattacharya\altaffilmark{M2},
I.~A.~Bond\altaffilmark{M3},   
M.~Freeman\altaffilmark{M4},
A.~Fukui\altaffilmark{M6},
Y.~Hirao\altaffilmark{M1},
Y.~Itow\altaffilmark{M5},
N.~Koshimoto\altaffilmark{M1},
C.~H.~Ling\altaffilmark{M3},
K.~Masuda\altaffilmark{M5},
Y.~Matsubara\altaffilmark{M5},
Y.~Muraki\altaffilmark{M5},
M.~Nagakane\altaffilmark{M1},
K.~Ohnishi\altaffilmark{M7},
N.~Rattenbury\altaffilmark{M4},
To.~Saito\altaffilmark{M8},
D.~J.~Sullivan\altaffilmark{M9},
T.~Sumi\altaffilmark{M1},
D.~Suzuki\altaffilmark{M2},
P.~J.~Tristram\altaffilmark{M10},
A.~Yonehara\altaffilmark{M11}\\
(The MOA Collaboration$^{M}$)\\
        and\\
M.~Dominik\altaffilmark{D1,R},    
U.~G.~J{\o}rgensen\altaffilmark{D2},
V.~Bozza\altaffilmark{D3,D4},
K.~Harps{\o}e\altaffilmark{D2},
M.~Hundertmark\altaffilmark{D1,D2},
J.~Skottfelt\altaffilmark{D2,D5}\\
(The MiNDSTEp Collaboration)
}
\email{\myemail}

\altaffiltext{$\ast$}{Based on observations obtained with the 1.3-m Warsaw telescope at the Las Campanas Observatory operated by the Carnegie Institution of Washington.}
\altaffiltext{O1}{Warsaw University Observatory, Al. Ujazdowskie 4, 00-478 Warszawa, Poland}
\altaffiltext{O2}{Department of Astronomy, Ohio State University, 140 W. 18th Ave., Columbus, OH 43210, USA}
\altaffiltext{O3}{Universidad de Concepci\'on, Departamento de Astronomia, Casilla 160-C, Concepci\'on, Chile}
\altaffiltext{O4}{Department of Physics, University of Warwick, Gibbet Hill Road, Coventry, CV4 7AL, UK}
\altaffiltext{M1}{Department of Earth and Space Science, Graduate School of Science, Osaka University, Toyonaka, Osaka 560-0043, Japan}
\altaffiltext{M2}{Department of Physics, University of Notre Dame, Notre Dame, IN 46556, USA} 
\altaffiltext{M3}{Institute of Information and Mathematical Sciences, Massey University, Private Bag 102-904, North Shore Mail Centre, Auckland, New Zealand} 
\altaffiltext{M4}{Department of Physics, University of Auckland, Private Bag 92019, Auckland, New Zealand} 
\altaffiltext{M5}{Solar-Terrestrial Environment Laboratory, Nagoya University, Nagoya 464-8601, Japan} 
\altaffiltext{M6}{Okayama Astrophysical Observatory, National Astronomical Observatory of Japan, 3037-5 Honjo, Kamogata, Asakuchi, Okayama 719-0232, Japan}
\altaffiltext{M7}{Nagano National College of Technology, Nagano 381-8550, Japan}
\altaffiltext{M8}{Tokyo Metropolitan College of Aeronautics, Tokyo 116-8523, Japan}
\altaffiltext{M9}{School of Chemical and Physical Sciences, Victoria University, Wellington, New Zealand}
\altaffiltext{M10}{Mt. John University Observatory, P.O. Box 56, Lake Tekapo 8770, New Zealand}
\altaffiltext{M11}{Department of Physics, Faculty of Science, Kyoto Sangyo University, 603-8555 Kyoto, Japan}
\altaffiltext{D1}{SUPA, University of St Andrews, School of Physics \& Astronomy, North Haugh, St Andrews, KY16 9SS, United Kingdom}
\altaffiltext{D2}{Niels Bohr Institute \& Centre for Star and Planet Formation, University of Copenhagen, {\O}stervoldgade 5, 1350 Copenhagen K, Denmark}
\altaffiltext{D3}{Dipartimento di Fisica ``E.R.\ Caianiello'', Universit\`{a} di Salerno, Via Giovanni Paolo II 132, 84084, Fisciano (SA), Italy}
\altaffiltext{D4}{Istituto Nazionale di Fisica Nucleare, Sezione di Napoli, 80126 Napoli, Italy}
\altaffiltext{D5}{Centre for Electronic Imaging, Dept.\ of Physical Sciences, The Open University, Milton Keynes MK7 6AA, United Kingdom}
\altaffiltext{O}{Optical Gravitational Lens Experiment (OGLE)}
\altaffiltext{M}{Microlensing Observations in Astrophysics (MOA) Collaboration}
\altaffiltext{R}{Royal Society University Research Fellow}

\begin{abstract}
We present the discovery of a Neptune-mass planet orbiting
a $0.8 \pm 0.3 M_\odot$ star in the Galactic bulge. The planet manifested itself
during the microlensing event MOA 2011-BLG-028/OGLE-2011-BLG-0203
as a low-mass companion to the lens star. The analysis of the light curve
provides the measurement of the mass ratio: $(1.2 \pm 0.2) \times 10^{-4}$, 
which indicates the mass of the planet to be 12--60 Earth masses. 
The lensing system is located at $7.3\pm0.7$ kpc away from the Earth 
near the direction to Baade's Window. 
The projected separation of the planet, at the time of the
microlensing event, was 3.1--5.2 AU.
Although the ``microlens parallax'' effect is not detected in the 
light curve of this event, preventing the actual mass measurement, 
the uncertainties of mass and distance
estimation are narrowed by the measurement of the source
star proper motion on the OGLE-III images spanning eight years, 
and by the low amount of blended light seen, proving that 
the host star cannot be too bright and massive.
We also discuss the inclusion of undetected parallax and 
orbital motion effects into the models, and their influence
onto the final physical parameters estimates.
\end{abstract}

\keywords{gravitational lensing: micro -- planetary systems}

\section{Introduction}
\label{sec:intro}

Both, the extrasolar planet distribution and the planet formation 
mechanisms are of a great interest in current astrophysics.
In order to gain insight into these matters, all possible methods
of planet detection should be exercised, since every method has its
own strengths, biases, and probes particular subspace of
the planetary system parameters.

The core-accretion theory of planet formation \citep{laughlin04,ida05}
predicts that the giant planets and Neptune-mass planets 
form beyond the ``snow-line'' of its hosts, where the solid
material density is greatly increased by frosting.
Although transit and radial-velocity methods of planet detection
find a number of giant extrasolar planets, these are mainly 
hot-Jupiters that migrated from the place of their formation.
Sensitivity of the mentioned methods to planets with orbits of a few AU,
and larger, is very limited -- typical sensitivity ends below 2.5 AU
\citep{johnson10}. For example, \citet{cumming08} studied periods of 2--2000 days with RV method, 
which corresponds to the mean semi-major axis of 0.31AU,
while the position of the ``snow-line'' can be approximated
with $\sim 2.7 M/M_\odot$ \citep{kennedykenyon08}.

The microlensing method is best suited for probing the planet population
beyond the ``snow-line'' being sensitive to gas giant planets, as well
as, Neptune-mass planets in the region of their formation \citep[see][for a review]{gaudi12}.
In the Galactic-scale lensing event the light from the distant 
star, bent by the gravity of a stellar-mass lens, typically passes
2--4 AU from this lens. If a planetary-mass companion to the lens is present
at these separations, 
it can disturb the image of the distant star, change its magnification, 
and therefore, manifest its own presence to the careful observer.

The lensing action does not depend on the light of the host star nor the planet,
neither their radii, and is only a weak function of their mass ($\sim \sqrt{M}$).
This gives the microlensing method great advantage in discovering cold planets
around all types of stars.

To date, more than $40\%$ of planets found by microlensing can
be classified as cold-Neptunes or sub-Saturns, 
and $\sim 30\%$ as giant planets.
Initial studies of the sample of microlensing planets show that 
$38_{-22}^{+31}\%$ of stars host cold super-Earths or Neptunes 
with separations in 1.6-4.3 AU range \citep{gould06}, and that 
they are $7_{-3}^{+6}$ times more common
than cold Jupiters \citep{sumi10}.

Although, the {\it Kepler} mission provided the evidence that 
the Neptune-mass planets are common on shorter-period orbits,
the gravitational lensing results strongly suggest that 
this is also the case for longer-period orbits.
Furthermore, studies of the planetary mass function based 
on microlensing sample \citep{gould10,cassan12}
confirm the increased abundance of planetary companions 
beyond the ``snow-line'' (see Fig.~8 of \citealt{gaudi12}),
where they are expected to form efficiently.

Despite only about 30 planets being known from the microlensing
technique to date, it has already proved itself as complementary 
to other methods of planet discovery and provided useful insights
into planetary population in the Galaxy \citep{gaudi12}. 
It is crucial, however, to further
work on expanding the sample of microlensing planet to facilitate 
these studies.
On one hand, the statistical strength of the arguments should be 
greatly improved,
on the other, the interesting edge cases are being found in
the process, for instance, planets in binary stars systems \citep[eg.][]{poleski14a,gould14,udalski15b},
or giant planets around low-mass stars \citep[eg.][]{poleski14b,koshimoto14,fukui15,skowron15}.
Together, with the sound statistics for more typical planetary systems,
it will be possible to predict the underlying frequency of these unusual systems.

Here, we report the discovery of $\sim 30 M_\earth$ planet (\object{MOA-2011-BLG-028Lb})
most likely located in the Galactic bulge that orbits moderately-massive
star ($\sim 0.75\,M_\odot$) at $\sim 3-5$ AU.
Due to the distant position of the planetary system in the Galaxy
(small parallax) and low-magnification nature of the event
(high uncertainties in the flux estimations), its
physical parameters could not be accurately derived from the light curve.
Nevertheless, we provide best estimations based on our understanding
of the Galaxy, the measured source star proper motion (from 8-years of 
the OGLE monitoring) and upper limits on the lens flux from the OGLE light curve.
We hope the future high-resolution imaging could provide additional
constraints.

In Section~\ref{sec:data} we describe the photometric observation
and the light-curves from each dataset taken into account. 
Section~\ref{sec:lc} presents microlensing model parametrization
and discusses additional effects taken into account when
modeling the light-curve data. 
The physical parameters of the system are derived in 
Section~\ref{sec:phys} with use of some additional inputs 
and assumptions about the Galaxy and the source star.
Results are presented in Section~\ref{sec:results} 
together with the discussion of future follow-up observations. 
We conclude the paper with Section~\ref{sec:conclusions}.

\section{Photometric observations and light-curve data}
\label{sec:data}

%
%
%
In March of 2011 the Microlensing Observations
in Astrophysics group (MOA) announced the candidate microlensing event:
MOA 2011-BLG-028\footnote{\url{https://it019909.massey.ac.nz/moa/{\allowbreak}alert/display.php?id=gb13-R-3-8819}}
on their {\it Microlensing Alerts}
webpage\footnote{\url{https://it019909.massey.ac.nz/moa/{\allowbreak}alert/alert2011.html}}. 
The event was also monitored by 
the fourth phase of the Optical Gravitational Lensing Experiment (OGLE-IV, \citealt{udalski15a})
and subsequently announced by the {\it Early Warning System}
(EWS)\footnote{\url{http://ogle.astrouw.edu.pl/ogle4/{\allowbreak}ews/ews.html}}
as OGLE-2011-BLG-0203\footnote{\url{http://ogle.astrouw.edu.pl/ogle4/ews/{\allowbreak}2011/blg-0203.html}}
in the batch of 431 microlensing events initializing the EWS at the OGLE-IV phase.

The Event's coordinates are
$(\alpha,\delta)_{\rm J2000} =(18^{\rm h}03^{\rm m}24.96^{\rm s},
-29^\circ 12' 48''\hspace{-2pt} .3)$ in equatorial coordinates and
$(l,b)=(1.7^\circ,-3.5^\circ)$ in Galactic  coordinates.
(The accuracy of the absolute position is of the order of 0.1 arcsec.)

The object was magnified from December 2010 until September 2011, 
and the magnification peaked on April 22nd, 2011 (HJD' = HJD - 2450000 = 5674) 0.4 mag brighter 
than the baseline level of $I=15.3$. 
Twenty days after the peak, on May 12--14 (HJD'=5694.2--5696.7), the short-time planetary 
anomaly was recorded by OGLE, MOA and Danish telescopes. 
The anomaly was spotted a couple of days after it was already finished, 
and the event followed a typical Paczy{\'n}ski light curve \citep{paczynski86} after that.
No other observatories managed to gather additional data on the planetary signal in
the event's light curve.

\begin{figure*}[ht]
\epsscale{1.11}
\plottwo{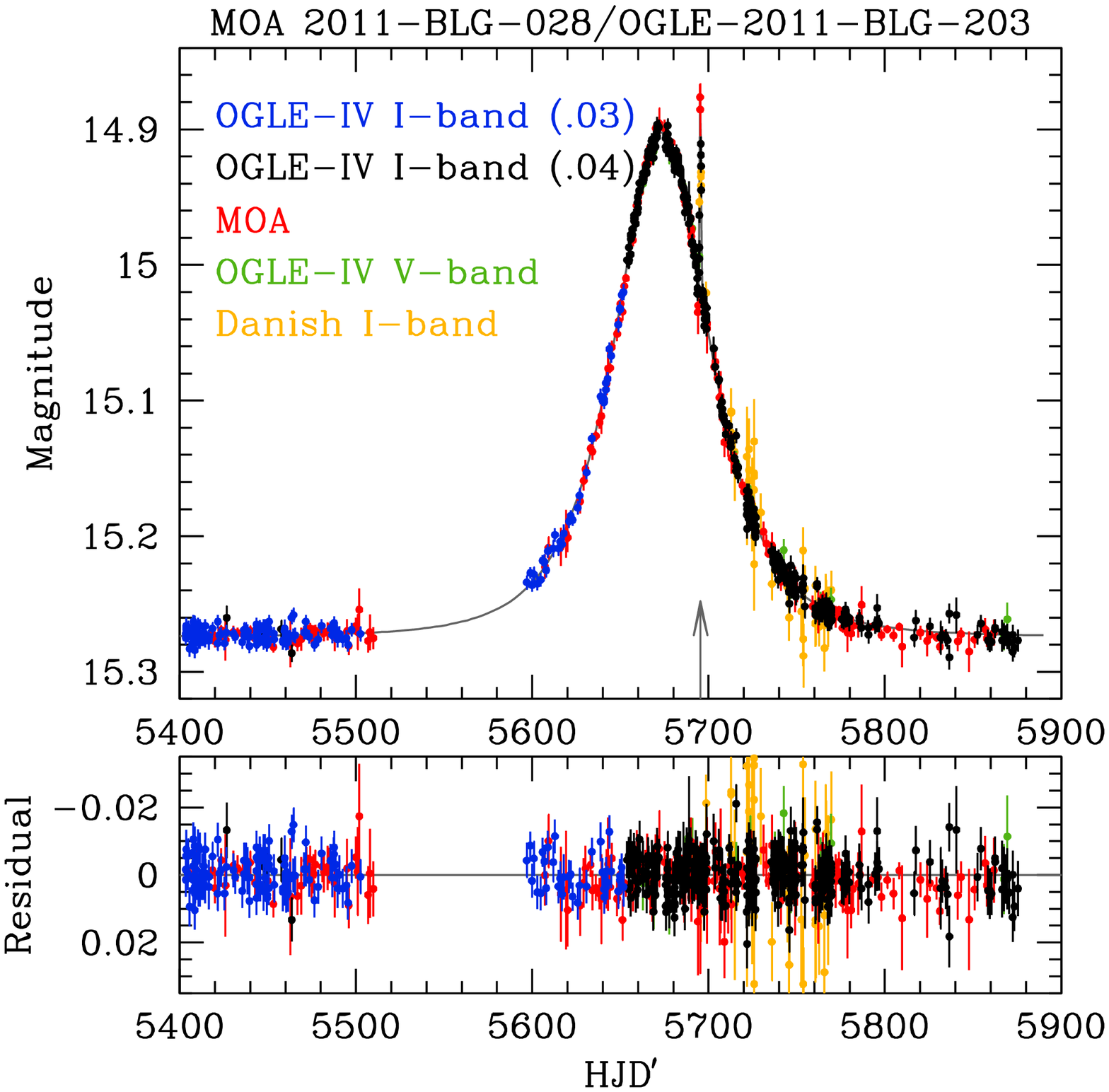}{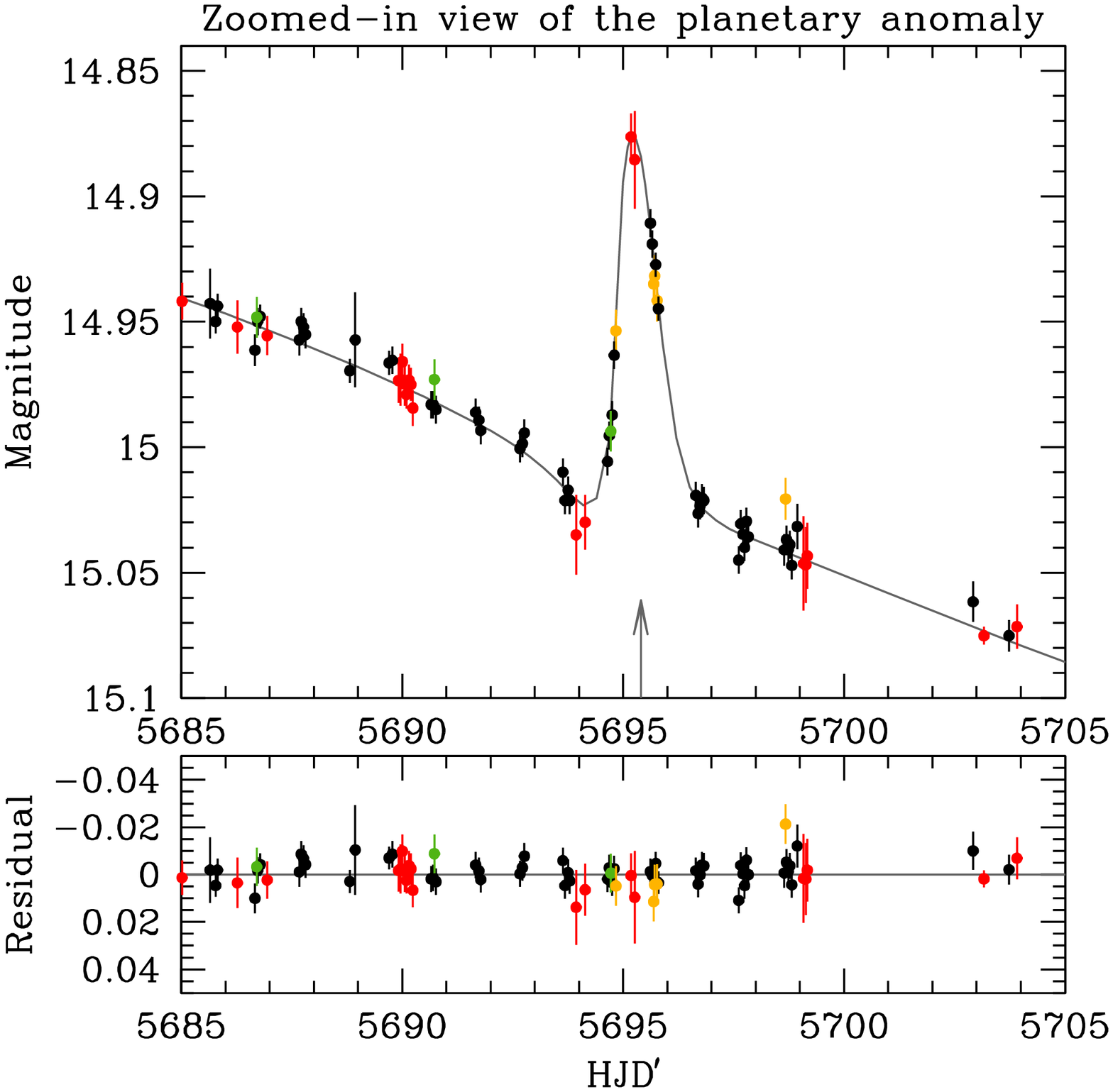}
\epsscale{1.0}
\caption{\label{fig:lc}
Left panel: 1.5-years long section of the light curve of the MOA 2011-BLG-028/OGLE-2011-BLG-0203.
Right panel: 20-days section centered around the planetary anomaly (marked with an upward arrow). 
The whole OGLE light curve for this object spans 15 years.
Only subset of data taken during the event that have been used in the final modeling are shown. 
Black line marks best-fit microlensing model where the light of a Galactic Bulge giant is magnified
for $\sim 200$ days (around 22nd April 2011) by a stellar object near light's path, and
further is disturbed for $\sim 2$ days (around 13th May 2011) by a low-mass companion to that object. 
Five datasets are presented as color dots together  with error bars used in the fitting process. 
Magnitude scale is calibrated to the OGLE-III photometric map.
The light curve of the second solution ($u_0<0$) is very similar and indistinguishable by eye on this plot, hence, 
only one solution ($u_0>0$) is presented.
Bottom part of each plot shows residuals against the best-fit model. (HJD'=HJD-2450000). 
OGLE-IV I-band light curve is split into the measurements made with the CCD 
detectors no. \emph{03} and \emph{04} of the OGLE-IV camera. Up to HJD' = 5650 majority of the measurements are done 
with the detector no. \emph{03}, while after that date, due to the slight change in the pointing model of the Telescope, 
all measurements, including the peak of the event and planetary anomaly, have fallen onto the detector no. \emph{04}.
}
\end{figure*}

Figure~\ref{fig:lc} shows the 1.5 year-long section of the light curve
covering the microlensing event and the planetary anomaly. This object 
was monitored by OGLE from 2001 and by MOA from 2006 and does not show any other variability 
outside of the period shown.

The OGLE survey uses the dedicated 1.3-m Warsaw Telescope located at Las Campanas Observatory in Chile. 
The MOA group observes with the 1.8-m telescope at Mt. John University Observatory in New Zealand.
The 1.5-m Danish telescope at ESO La Silla in Chile is operated by 
the MiNDSTEp Consortium\footnote{\url{http://www.mindstep-science.org/{\allowbreak}about\_us.2011.html}}.

While the microlensing survey groups (OGLE and MOA) constantly monitor 
the relevant regions of the Milky Way in order to identify and characterize microlensing events, 
the MiNDSTEp Consortium, among other {\it follow-up teams}, monitors 
only promising microlensing events in effort of detecting extrasolar planets \citep{dominik10}.
It happened that the automatic prioritization algorithm, used by this group, fortunately flagged the MOA 2011-BLG-028
as a potentially interesting event to observe right before the actual planetary anomaly occurred.
The first data point from the Danish telescope considered in this work
is, actually already, during the planetary anomaly (at HJD'=5694.84).

In this work, we also use data from the previous phase of the OGLE survey
(OGLE-III) which was operating from 2001 to 2009 at
1.3-m Warsaw telescope. The {\it V}- and {\it I}-band data
comes from the project's final data reductions \citep{udalski08}.
The calibrated data on stars in the neighborhood of the event
are taken from the Galactic bulge photometric maps
\citep{szymanski11}, thus, all OGLE magnitudes reported
in this paper are standard {\it V} (Johnson) and {\it I} (Cousins) magnitudes.

\subsection{Data Preparation}

In the light curves from both microlensing surveys the main event is clearly 
detected and shows all prerequisites of the microlensing 
event by a stellar system. The planetary anomaly is detected by three telescopes
and its shape is what we expect from the added 
magnification of the major image by the planetary companion \citep[eg.][]{mao91}.

We expect that the projected position of the planetary system on the plane of
the sky can be modified during the curse of the event ($\sim 200$ days)
by its orbital motion. Also, the motion of the observer on Earth's orbit
can modify the geometry of the event via, so called, ``microlens parallax'' effect.

The decade-long light curve of the stars involved in the event
show no signs of periodic or non-periodic variability nor transient 
outbursts. 
Since we expect the microlensing to be the source of the detected
magnification, we fit the microlensing model to all data sets
and require all data sets yield $\chi^2$ per degree of freedom equal to unity
-- this is done by rescaling the uncertainties.
We use standard approach to the measurement errors rescaling, as described for
instance by \citet[Sec~2.2]{skowron11}. 

Theoretically founded expectations about the shape of the light curve
in conjunction with the great redundancy of the data allows us to 
easily judge which data points can be classified out-right as outliers.
Also, it is possible to identify suspect spans of the light curve data of
being taken under sub-optimal conditions for further evaluation and
possible removal from the final dataset. Since, beside the
short-term planetary anomaly, we are seeking for slow-evolving trends
in the light curve, characteristic for the parallax effect or lens system 
orbital motion, we are conservative and remove all suspect data.

\subsection{OGLE-IV light curve data}
\begin{figure}[ht]
\epsscale{0.8}
\plotone{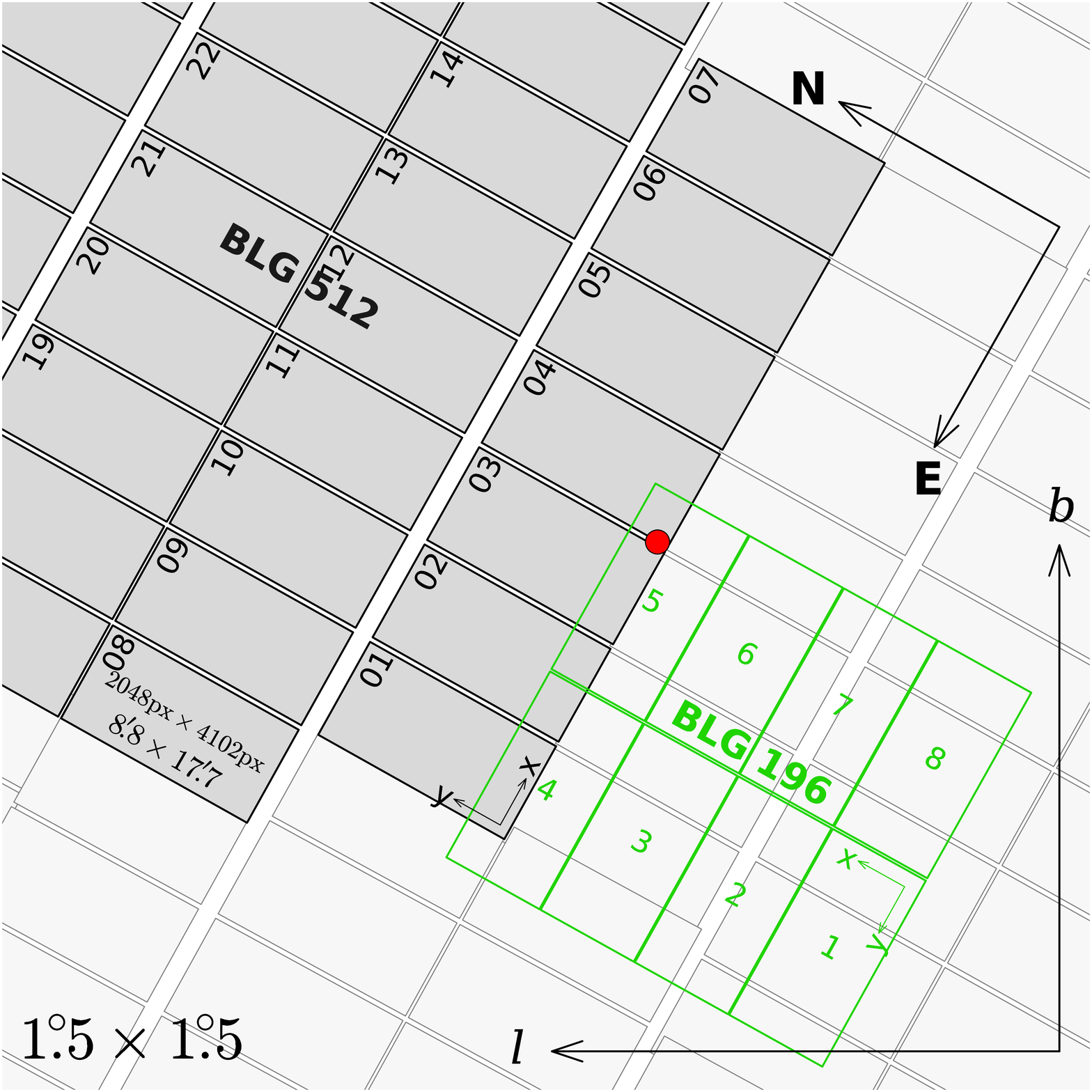}
\epsscale{1.0}
\caption{\label{fig:ccds}
The location of the microlensing event MOA-2011-BLG-028/OGLE-2011-BLG-0203 in respect to the OGLE-IV field \emph{BLG512} and the OGLE-III field \emph{BLG196}. 
The red circle has radius of 1 arc minute and marks the position of the event. 
In the OGLE-IV survey, the event falls into the gap between the CCD detectors number \emph{03} and \emph{04}, close to the corners of both detectors.
Due to the scatter in the telescope pointing, 
the microlensing event was registered on both CCD detectors. It was also automatically 
discovered by the Early Warning System. The measurements, however, are split 
into two separate light curves, for both detectors, and require additional 
cross-calibration.
Earlier, during the course of the OGLE-III survey, the region of the future microlensing event was monitored for 8 years with the CCD detector number \emph{5} in the field \emph{BLG196}.
Galactic north is up and Galactic east is to the left. The plot has $1^\circ\hspace{-2pt}{.5} \times\,1^\circ\hspace{-2pt}{.5}$ field of view. Each CCD detector covers $8{^\prime}\hspace{-2pt}.8 \times\,17{^\prime}\hspace{-2pt}.7$ of the sky.
}
\end{figure}

The object described in this paper falls into the gap between two CCD
detectors of the OGLE-IV camera in the standard field
 no. \emph{BLG512} of the Galactic bulge survey. 
It is also located close to the corners of both detectors. 
See Fig.~\ref{fig:ccds} for the detailed view.
Fortunately, the typical pointing scatter of the telescope (rms $\sim 60 px$) causes
some number of measurements to be done with the detector no. \emph{03} 
and some with the detector no. \emph{04}. 
The original OGLE-IV \emph{BLG512} pointing in 2010 caused the star 
to fall more often in the \emph{03} detector. However, after early 2011 adjustments 
of the position of the OGLE-IV high cadence fields, including \emph{BLG512},
the majority of observations of the star was recorded on the \emph{04}
detector. Finally, a small, temporary adjustment to the \emph{BLG512} field
(about 50 pixels) was additionally introduced after the discovery of the planetary
anomaly to secure good coverage of the late stages of the microlensing event.
Since this field is monitored with a high cadence of up to 10--30 exposures
per night, the two datasets combined yield good, continuous coverage
of the event throughout the season, albeit with a lower cadence of $\sim 2$ per night.

\notetoeditor{(fig:ccds) black and white for print, in color only in the electronic version}

The routine OGLE-IV calibrations to the Johnson-Cousins photometric system do
not perform well for this object due to its extreme position (at the edge and near 
the corner of the detector). 
Fortunately, this region of the sky was also densely monitored by the previous
phase of the OGLE survey (OGLE-III, \cf\ Fig.~\ref{fig:ccds}). Where, the object in question was
well measured with over 1300 individual observation during eight years of the project
and its calibrated {\it I}- and {\it V}-band magnitudes are given by
\citet{szymanski11}: $(V-I, I) = (1.829, 15.275) \pm (0.011, 0.006)_{\rm stat}$.

\begin{figure}[ht]
\plotone{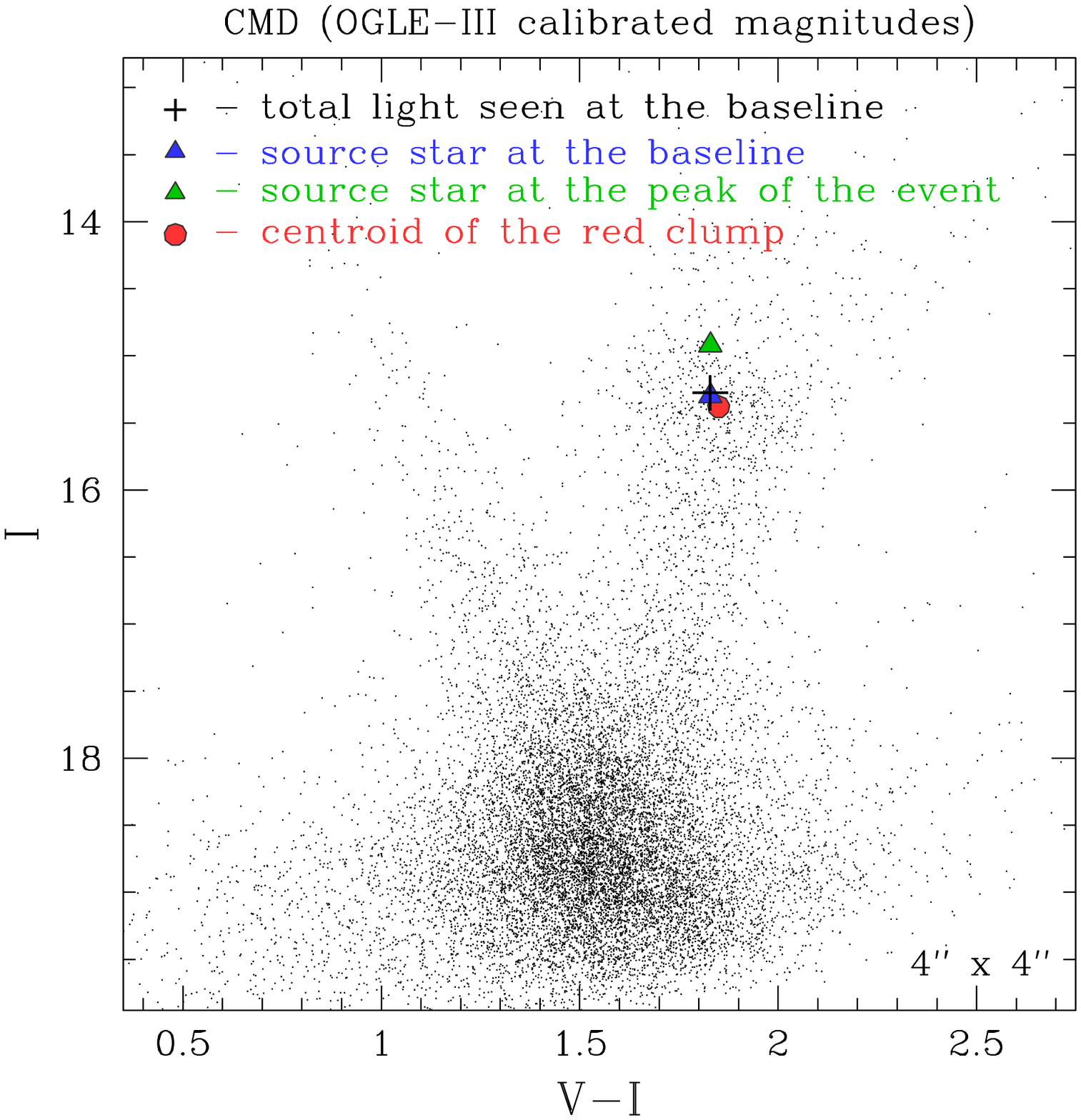}
\caption{\label{fig:cmd}
The Color-Magnitude Diagram (CMD) based on the OGLE-III photometric map \citep{szymanski11} of the $4'\times 4'$ region around the source star.
The center of the red clump giant's region is marked with red circle.
The position of the studied object (at baseline) is marked with a blue triangle.
Microlensing model strongly favors solutions with small amounts of additional
light (blended light, see Table \ref{table:ulens}).
Therefore, the light we see at the baseline is mainly the light of the microlensed star. 
We infer, from its position on the CMD that it is, most probably, a red clump giant in the Galactic bulge.
}
\end{figure}

The magnified source star is most likely a red clump giant in the Galactic 
bulge (see Fig.~\ref{fig:cmd}) and account for bulk of the light seen 
before and after the event (the microlensing models with a very
low amounts of not-magnified light are strongly preferred
by the light curve  -- see. Section~\ref{sec:lc}). 
Hence, we can calibrate the whole OGLE-IV light curve to the OGLE-III
Johnson-Cousins magnitudes, as measured before the event, and
not worry about the color changes during the event, and thus,
different color terms. Any potential errors this procedure introduces 
are insignificant compared to the OGLE-III calibration uncertainties 
(0.01--0.02 mag).

\notetoeditor{(fig:cmd) black and white for print, in color only in the electronic version}

In order to accurately measure the {\it I}-band brightness evolution during the event,
we construct a custom template image by averaging a dozen of science frames 
taken under good weather conditions and where the object position fallen
not less then 0.5 arc minute from the edge of the detector. The centroid of the 
source star is calculated from 7 frames in April 2011 when the magnification
was the highest. Then, we perform an optimized photometry with the OGLE-IV pipeline
(based on the Difference Image Analysis, DIA, \citealt{wozniak00}) 
measuring the variable light of the event
with profile photometry on the subtracted image at the fixed position of the source.
The resulting OGLE-IV light curve consists of 1974 measurements with the CCD
detector no. \emph{04} and 275 measurements with the detector no. \emph{03}, 
out of which we remove one data point and 25 data points, respectively, 
as outliers or data taken under sub-optimal conditions. 
We rescale the errorbars by adding in quadrature 3 mmag and 3.5 mmag,
respectively.

The {\it V}-band observation are reduced with the standard image
 subtraction pipelines, as described by \citet{udalski03},
 from the measurements made with the CCD detector no. \emph{04}.
There are 63 measurements after one outlier is removed, 
and the errorbars are adjusted by adding 6 mmag scatter in quadrature.

\subsection{MOA light curve data}
The MOA  data were reduced using the standard image
subtraction pipeline used be the survey group and described by \cite{bond01}.
We fit for possible airmass and seeing correlations
and correct the standard photometry accordingly.

The MOA light curve consists of 7183 measurements, from which we take only 
5904 with the reported uncertainties lower than 0.01 mag.
The typical sampling rate during the observing season was $\sim 5$ per night.
Due to the season-to-season low amplitude shifts ($\sim 0.01$ mag) present in the light curve
-- most likely an instrument-introduced systematic effect -- we
remove all data outside of the 2011 season in order to avoid spurious 
signals in the microlens parallax measurement.
We measured that different baseline levels between 2010 and 2011 seasons were introducing
asymmetricity to the light curve that resulted in the apparent 2-$\sigma$
detection of the parallax signal. This signal vanished when the 
MOA baseline data were cropped.

We also skip observations at the very beginning and at the end of the 
season, since these are not crucial for characterizing the event
(due to the existence of other data), but
carry some risk of introducing additional systematic errors. The MOA telescope  
on-average observes worse weather than the Chilean sites and the high airmass 
and high effective seeing near the seasonal break,
in conjunction with the large pixel size and significant crowding 
toward this pointing, makes the measurements more 
challenging. 

From the HJD' range of 5620 -- 5845 we 
take 784 data points and we bin them in one-day intervals.
We leave the period $\pm 5$ days around the planetary anomaly not binned,
since the variations of the light curve in this region 
have shorter time-scale.
Such constructed light curve consists of 131 data points, 
and the errorbar scaling factor used is 2.0.

\subsection{1.5-m Danish telescope light curve data}
The follow-up monitoring of the MOA 2011-BLG-028 event by the Danish telescope at 
the ESO Observatory, La Silla, Chile
started before the planetary anomaly happened, and was the result of strategy
described by \citet{dominik10}, in which, large fraction of promising 
microlensing events discovered by the survey groups are 
monitored with moderate cadence in the anticipation that one of them would
unveil the existence of a planetary companion. This serendipitously happened
for this event.

Unfortunately, the event took place at the same time as the initial tests 
of the newly installed lucky imaging (LI) camera. 
Therefore half of the observations were done with the old conventional 
CCD camera and half with the new EMCCD (LI) camera. 
Recent observations have demonstrated the ability to obtain as accurate 
photometry with the LI technique as with conventional CCDs at the same 
time as benefiting from the high speed and increased spatial resolution 
of the LI technique \citep{harpsoe12,skottfelt15},
and the LI camera is now the standard instrument during the MiNDSTEp 
microlensing observations. The LI light curve started before the anomaly 
while first data from the standard CCD camera are at the rising part 
of the anomaly (as shown on the second panel of Fig.~\ref{fig:lc}). 
The mixed approach lowered the cadence of each data set, and because 
of the difficulties in cross-calibrating both techniques from this early 
testing phase, we must treat the two light curves as separate data sets. 
The LI light curve
shows significantly larger scatter
then the standard one, and we do not use it in the modeling. 

Albeit lower cadence, the standard {\it I}-band light curve is still useful to further 
confirm the amplitude and timing of the anomaly, as the Danish telescope 
is the third telescope to have observed it.
Out of 57 data points, we remove 7 as clear outliers and increase errorbars
of the remaining 50 data points by a factor of 2.55 to ensure $\chi^2$ per
degree of freedom $\sim 1$.


\section{The Light Curve Modeling}
\label{sec:lc}

\subsection{Microlensing equations and parameters}
\label{sec:eqs}
A microlensing event is a transient magnification 
of a light coming from the distance star ({\it source} of light: S)
by a massive object passing near the line of sight (the {\it lens}: L). 
Magnification is a direct result of stretching, bending and
increasing the number of source-star's images on the sky by gravitational 
influence of the lens object. 

\begin{figure*}[h]
\epsscale{1.1}
\plottwo{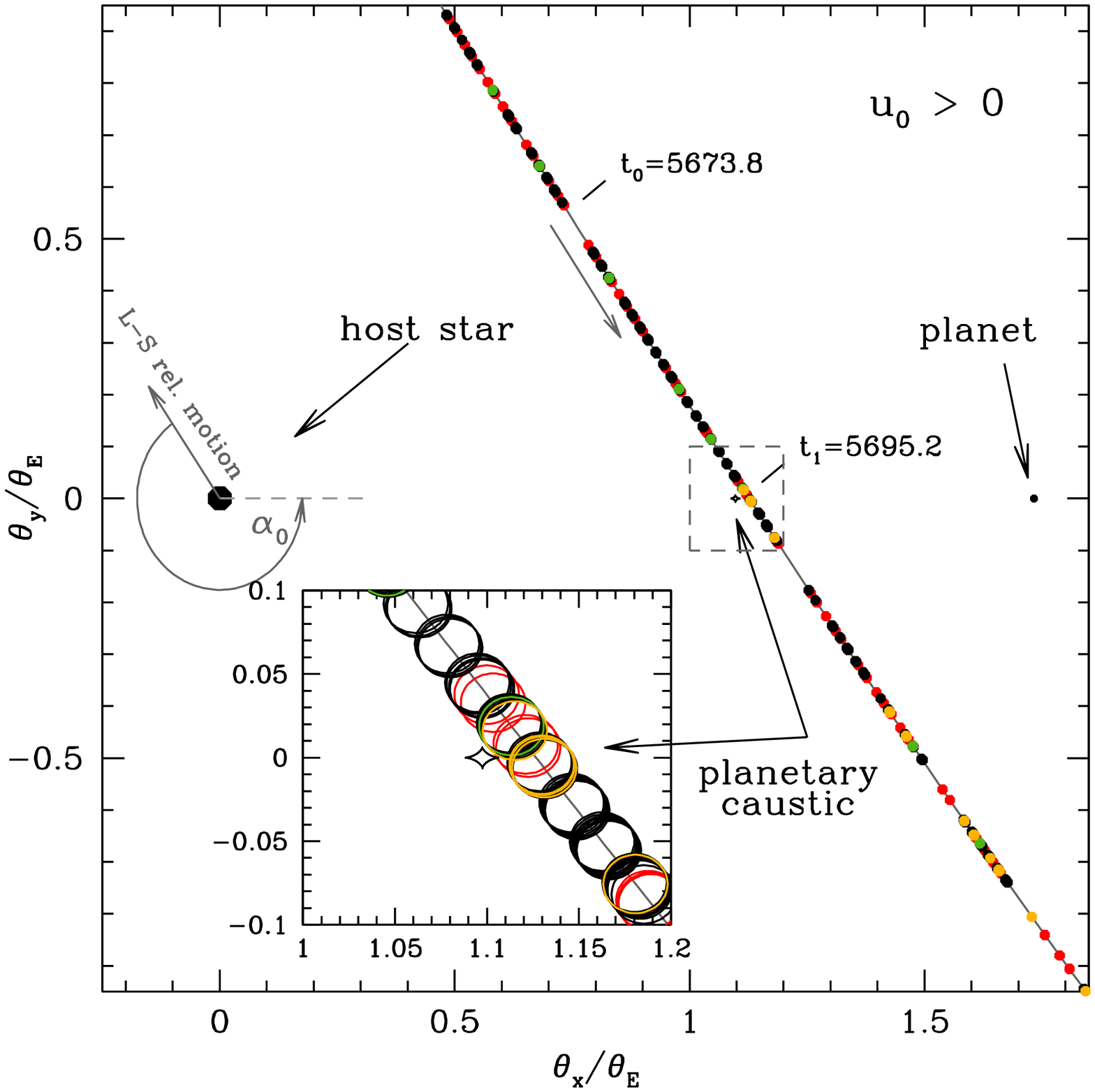}{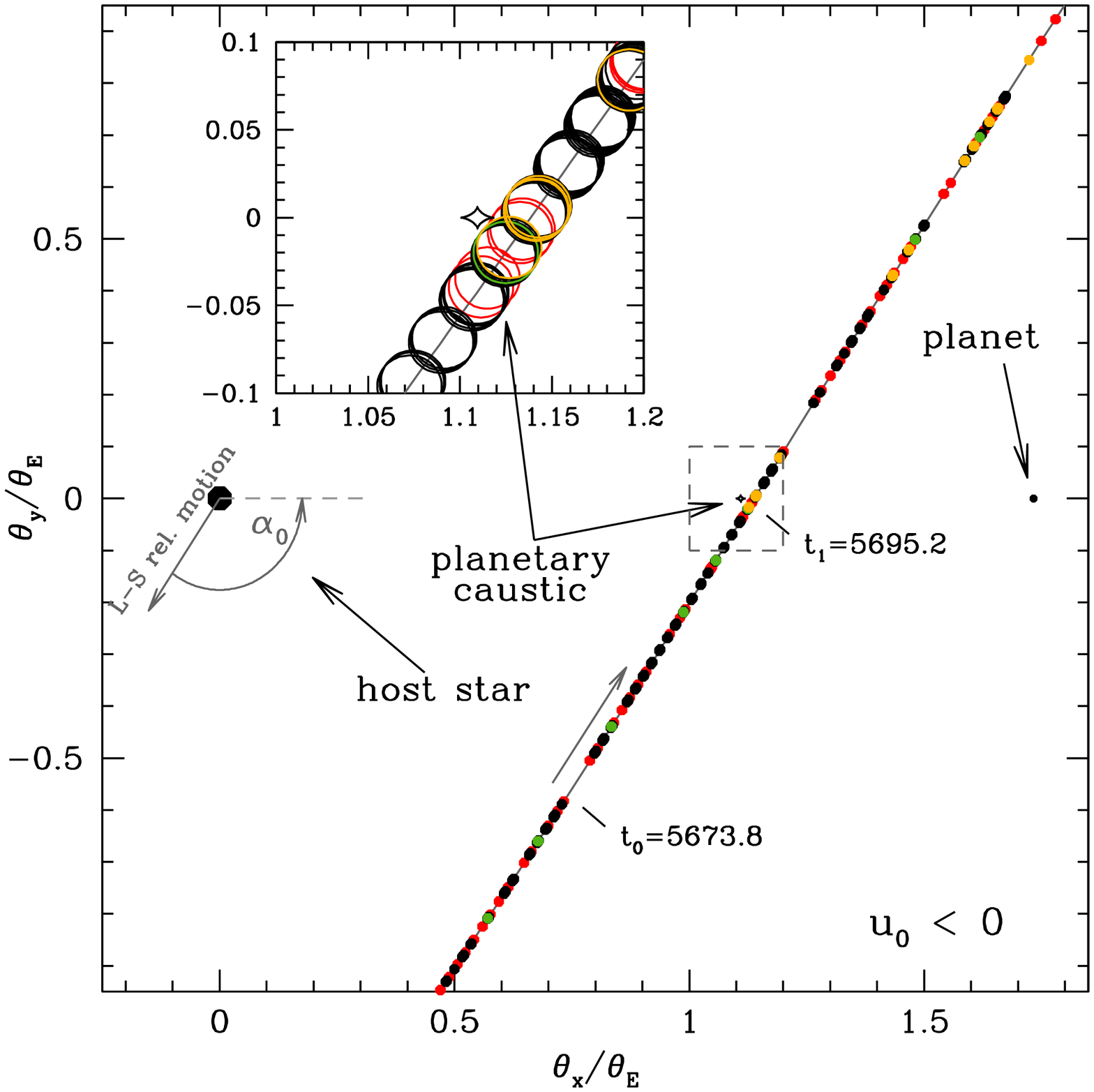}
\epsscale{1.0}
\caption{\label{fig:geom}
The lens geometry and the source trajectory behind the lens projected
onto the plane of the sky in the units of angular
Einstein radius ($\theta_\e$). 
Left (right) panel show $u_0>0$ ($u_0<0$) solution. 
The lens components: host and planet, by convention
are located along the $x$ axis.
Origin is at the Center of Mass of the planetary system.
Main panels show 70-days-long trajectory of the source and insets show
its 7-days-long segment around the second magnification maximum caused by the source passage close
to the planetary caustic.
Times of both magnification maxima  ($\sim5673.8$ and $\sim5695.2$) are indicated with $t_0$ and $t_1$.
The positions of the source center
are marked along the best-fit source trajectory (gray line)
for the epochs its brightness was measured by the telescopes.
Open circle show the projected source size ($\rho_*$) and position at the different epochs, 
and diamond-like shape represents the planetary caustic.
Central caustic is very small and close to the host star, hence, not visible in the plots.
Colors follow the convention from the Fig.~\ref{fig:lc}.}
\end{figure*}

In the case of perfect alignment
of the observer, the lens and the source, observer sees one image
in the shape of a ring around the lens -- called the Einstein ring -- 
and its angular radius on the sky ($\thetae$) depends on the distances 
($D_{\rm L}$ and $D_{\rm S}$) and the 
mass of the lens ($M_{\rm L}$) in the following way:
\begin{equation}
  \label{eq:thetae}
  \thetae = \sqrt{ \kappa M_{\rm L} ( {\rm AU}/D_{\rm L} - {\rm AU}/D_{\rm S} )  },
\end{equation}
where $\kappa = 8.144\,\mas\,{M_\odot}^{-1}$ \citep[cf.][]{gould00}.
Toward the Galactic bulge, most of the potential source stars are 
located at $D_S \sim 8\, \kpc$ and lens stars are typically at 
$D_L \sim 4-7\, \kpc$, hence, for stellar-mass lenses 
$\thetae \sim 0.5-1.0\,\mas$. 
Typically, the relative lens-source proper motion in the Galaxy is 
of the order of $\mu_{\rm rel} \sim 2-5\,\masyr$ (for disk-bulge 
lensing events it is $\sim 4\, \masyr$, and for bulge-bulge lensing events: $\sim 2.5\,\masyr$). 
Thus, the time for the source star to cross the Einstein ring 
of the lens is:
\begin{equation}
  t_\e = \frac{\thetae}{\mu_{\rm rel}} \sim 20-100\, {\rm days},
\end{equation}
a, so called, the Einstein ring crossing time 
or the ``Einstein time''.
The magnification during the microlensing events is, therefore, evolving on the
time-scale of days and months. It reaches its maximal value
at the time, $t_0$, when the projected distance between the lens and the source star
becomes minimal -- denoted with $u_0$ and expressed in the units of $\thetae$.
By convention, we reserve the positive (negative) values for the $u_0$
parameter for cases in which the lens is passing the source on its right (left).
See Figure~\ref{fig:geom} for reference. 

It is sometimes beneficial to introduce the value of $t_{\rm eff} = u_0 t_\e$ to
be used as a model parameter in the fitting process instead of $t_\e$ or $u_0$.


The three introduced parameters ($t_\e$, $t_0$ and $u_0$),
or equivalently ($t_\e$, $t_0$ and $t_{\rm eff}$), describe 
the time evolution of the magnification during the simplest 
microlensing event, where both, 
lens and source, are single stars and the projected 
relative motion can be approximated as rectilinear \citep{paczynski86}.

\vspace{3\parskip}

The existence of the planetary companion to the lens can
influence the magnification pattern in its vicinity, and therefore,
require an introduction of additional parameters to describe 
the observed light curve well. These are: $s_0$ -- the projected 
separation of the host star and its companion at some specified time ($t_{0, orb}$,
here fixed to 2455678.0), $q$ -- a mass ratio of the planet to 
its host, and $\alpha_0$ -- an angle of the sky-projected 
planet-host axis at the specified time $t_{0, orb}$, measured 
counter-clockwise from the direction of the lens-source relative proper
motion $\muvec_{\rm rel}$.

During the months magnification is observed, the relative position
of the host and the planet can be evolving due to their orbital motion.
We describe this effect with two constant rates of change: 
$ds/dt$ and $d\alpha/dt$, for the separation and the projected angle
of the planet-host axis, respectively. 
We choose the center of mass of the lensing system as a
reference point for $t_0$ and $u_0$, and for the center of the 
coordinates system on the plots.

The observer located on Earth is always experiencing acceleration
from the Sun. Thus, for a few-months-long microlensing event,
even with the absence of the observable acceleration 
of the source star or originating inside of
the lensing system,
the evolution of the lens-source projected position (as seen by the Earth observer)
might not be well approximated by the rectilinear motion.
We hence fix the velocity of the observer frame to the Earth's velocity at 
the fiducial time $t_{0,par} = 2455678.0$ and use 
the geocentric parallax formalism \citep{gould04} to describe the
observer deviations from this motion. To parametrize the influence of the
Earth motion on the event's configuration, we
take advantage of the microlens parallax vector ($\pivec_\e$), which
has the same direction as the lens-source relative motion, while its
magnitude is a ratio of the Astronomical Unit to the radius of
the Einstein ring projected from the source onto the Earth's orbit.
It is a useful parameter, as it ties the scale of the Earth's orbit to 
the scale of the Einstein ring.
The simple projection gives:
\begin{equation}
 \label{eq:pirel}
 {\rm AU}/D_{\rm L} - {\rm AU}/D_{\rm S} = \thetae \pi_\e.
\end{equation}

Note, that by combining Eqs~(\ref{eq:thetae}) and (\ref{eq:pirel}) we have:
\begin{equation}
 \label{eq:mass}
 M_{\rm L} = \thetae/\pi_\e/\kappa.
\end{equation}
Thus, the measurement of $\pi_\e = |\pivec_\e|$ and the angular scale of 
the Einstein ring ($\theta_\e$) immediately yields the mass of the lensing system
and, together with the estimation of the source's distance,
provides the distance to the lens (Eq.~\ref{eq:pirel}).

\subsection{Extended source star}
The planetary anomaly is a result of the source star passing close
to the planetary caustic (\cf\ Fig.~\ref{fig:geom}). Due to the finite angular size of the source,
the observed magnification pattern is smoothed out in time, while 
different parts of the star's disk are being strongly magnified 
by the caustic proximity. 
To quantify this effect, we use $\rho_*$ parameter, which is the 
radius of the source's disk in respect to the Einstein ring radius, 
or, equivalently, we use the time $t_*$, in which the source
star passes the distance of its angular radius ($\theta_*$). 
We have the following relations:
\begin{gather}
  \rho_* = \frac{\theta_*}{\theta_\e} = \frac{t_*}{t_\e} \\
  \theta_\e = \frac{\theta_*}{\rho_*} = \theta_* \frac{t_\e}{t_*}
  \label{eq:theta_rho}
\end{gather}
The measurement of this effect from the light curve,
while knowing the angular radius of the source (see Sec.~\ref{sec:thetastar}), 
allows us to measure the angular size of the Einstein ring ($\theta_\e$).

For the brightness profile of the disk of the source star,
we adopt the square-root limb darkening law and use coefficients 
provided by \citet{claret00} in Table 32 for $v_t = 2$, 
solar metallicity, $T_{\rm eff} = 4750\,{\rm K}$, and $\log g = 2.5$ 
-- as we find appropriate for the red clump giant in the 
Galactic bulge, for which we observe $(V-I)_{0} = 1.04$ 
(see Section~\ref{sec:thetastar} for the characterization of 
the source star):
\begin{align}
  c_{I\mathrm{-band}},\, d_{I\mathrm{-band}} &= 0.2530,\, 0.4713, \\
  c_{\mathrm{MOA-}R}, \, d_{\mathrm{MOA-}R}  &= 0.3017,\, 0.4443, \\
  c_{V\mathrm{-band}},\, d_{V\mathrm{-band}} &= 0.6035,\, 0.2386.
\end{align}
Coefficients for non-standard MOA-{\it R} filter are calculated as
linear combination of {\it R}-band and {\it I}-band with
30\% and 70\% weights.

\subsection{Basic model parameters from the light curve}
Since the coverage of the light curve is very dense, and the shape of the anomaly
is well defined, the microlensing nature of the event is clear and, furthermore, there have to be at least 
two bodies in the lensing system. The Einstein timescale of the main brightening is
$t_{\rm E} \approx 33$ days and the impact parameter is $u_0 \approx 0.9$. The half-duration time of the planetary
anomaly is $\sim 1$ day and its peak is $\delta t \sim 21$ days after the peak of the main feature.
If the source star radius was small in respect to the Einstein ring radius, 
the mass ratio we could guess from these values would be $\sim \sqrt{1/33}$ = $10^{-3}$, 
otherwise, it would be an upper limit. Thus, for a stellar-mass host star, the 
light curve immediately points to a planetary-mass companion.

One can estimate the separation of the planetary caustic from the host star as:
$u_c = \sqrt{t_{\rm eff}^2 + \delta t^2}/t_\e = \sqrt{ u_0^2 + (\delta t/t_\e)^2} \sim 1.1$,
and, since, the planetary anomaly resembles a major image perturbation, we can estimate
the planet-host separation from: $s - 1 / s = u_c$ to be $s \sim 1.7$.
Also, the angle of the binary axis in respect to the source-lens trajectory 
will be $\pm \arctan(t_{\rm eff}/\delta t) = \pm 55 \deg$. 
This leads to the angle of the binary axis in respect to the lens-source relative 
proper motion of $\alpha_0 \approx (180 \pm 55) \deg \equiv \pm 125 \deg$. 

The by-eye estimated values of the model parameters $(t_0, t_{\rm eff}, t_\e, q, s_0, \alpha_0)$ are (5674.0, 30, 33, $10^{-3}$, 1.7, $\pm125$), where times are in days and angle in degrees.

\subsection{Light curve modeling}
When starting from the estimated values for microlens model parameters it is straightforward to converge to 
the satisfactory fit to the observed light curve.

We test for other models, eg. with minor image perturbation, or with binary source, but
do not find appropriate explanation of features in the light curve with them.
There is, however, a symmetric solution with $u_0 < 0$ (i.e., with the lens passing the source
on its left), which is mathematically indistinguishable, from the $u_0 > 0$ solution, 
in the case of static binary lens \citep[c.f.][]{skowron11}.

We use $\chi^2$ as a goodness-of-fit measure and use Marcov Chain Monte Carlo (MCMC)
method for sampling the parameter space around the solutions and evaluation of the parameter 
uncertainties. Following \citet{skowron15}, we calculate microlensing magnifications 
with hexadecapole and inverse ray shooting algorithms, where we use \citet{skowron12lib} method
and code library \citep[described in][]{skowron12} for solving the lens equations.

The best-fit model light curve is plotted as a continuous line in Fig.~\ref{fig:lc}, and
geometry of the event is presented in Fig.~\ref{fig:geom}.

\subsection{Non-negative blended light}
\label{sec:posflux}
The term {\it blended light} reefers
to the additional light measured at the position of the microlensing event, 
that was not magnified during its progress. This might be the
result of the field crowding, where some unrelated star happen to lie inside 
the source star's seeing disk, or might come from the distant companion
to the source star or the lens, or might come from the lens itself, or any combination 
of the above. Therefore, the blended light we measure from the light curve is
an upper limit on the brightness of the lens (or any other object in the seeing disk)
-- we will make use of this fact later, in Section~\ref{sec:ib}.

On the other hand, the best-fit microlensing model reports negative blending
on a few percent level, 
i.e., the negative amount of additional light is preferred by the mathematical
model of the event. 
In other words, the minimal $\chi^2$ is obtained when the source star
is brighter than the observed baseline brightness at the position of the event.
Then, it is possible to obtain the observed maximal brightness at the event's peak
with the lower amplification provided by the microlensing model.
At the event's baseline, however, when the amplification of the source star is
by definition equal to 1, there needs to be some negative light added, to
recover lower level of observed baseline brightness.

The negative source of light is unphysical, and thus, we limit our solutions
to only those with positive blended light. This lower limit ($\equiv 0$) on blend's flux
is, at the same time, a lower limit on the peak magnification, so
the upper limit on the impact parameter ($|u_0|<0.94$).

Actually, when dealing with the real scientific data, in some fraction of 
observed microlensing events, a low amount of negative blended 
flux is expected. 
We recognize, that it is possible for the 
profile (PSF) photometry, which is performed on the {\it baseline object},
to overestimate the background level in the crowded fields of Galactic 
bulge. This, it turn, would underestimate the object's brightness
and act as a negative source of light. This effect, however,
cannot be significantly larger than the faintest objects measured 
in the field, which for the OGLE-IV is $\sim 21$ magnitudes, and thus,
could significantly impact only events with very faint sources. 
In the case of MOA 2011-BLG-028,
the source star is very bright -- a red clump giant -- and 21 magnitude
corresponds to only $0.5\%$ of the baseline flux.

We treat the solutions with negative blend flux as being
allowed solely by the mathematical description of the event,
but not as a physical possibility.

\subsection{The microlens parallax}
Best-fit models including the microlens parallax effect are located very close to
the $\pivec_\e \equiv {\bf 0}$ in respect to the uncertainties in $\pivec_\e$. 
This tells us that the microlensing parallax effect, 
i.e. observer's motion along the Earth's orbit,
 is not detected in the light curve. 
Actually, this is not surprising, as most of the lenses are located 
in the Galactic bulge \cite[cf.][]{dominik06},
and therefore, the Einstein radius projected from the source onto the Earth's orbit is
very large. This makes any shift of Earth position along the orbit very hard to notice in
the light curve, unless there are strong magnification features, and they
are observed throughout significant length of time (for instance, in the case of {\it resonant}
caustic, cf. \citealt{skowron15}).

One can assume that if the microlens parallax 
is not detected, we do not have any information about it.
Therefore, the only recourse is to relay on the statistical
expectations on the value of the parallax vector (based on the Galactic models)
in order to estimate 
the physical parameters of the system (for example \citealt{beaulieu06,koshimoto14}).
In reality, this should rarely be the case. 

While we still need to employ our
expectations about the lenses in the Galaxy, we can also 
use limits on the microlens parallax vector derived from the light curve.
In short, not all values of the parallax vector, that we would expect
to see from the lens population, are consistent with the
particular light curve. 

Typically, these limits are one-dimensional,
as it is more easy to measure $\pi_{\e,\|}$ \citep{gould94},
and they can still rule out noticeable portions of the parameter space,
therefore are useful and should be employed \cite[eg.][]{gould06}.

\begin{figure*}[ht]
\plotone{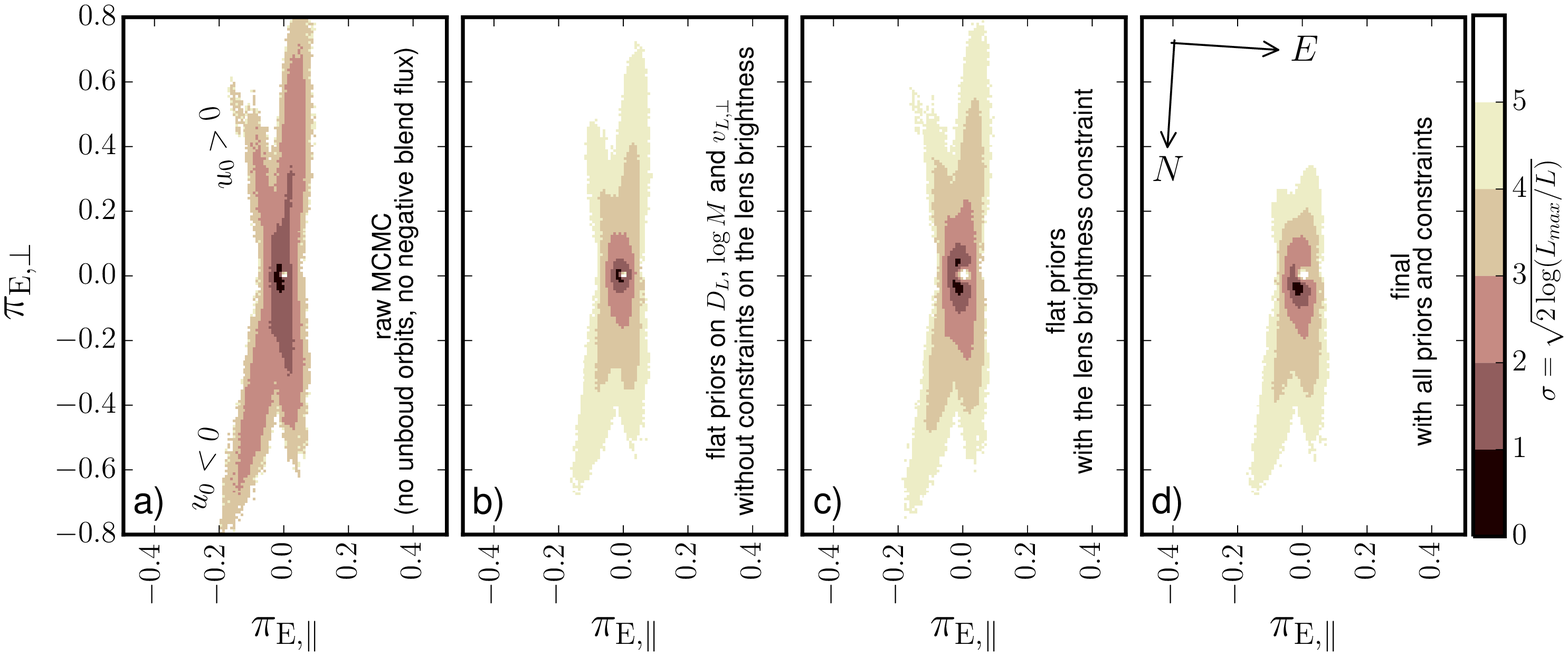}
\caption{\label{fig:prior}
Evolution of the parallax vector posteriors depending on the choice of priors.  
The $\pi_{{\rm E},\parallel}$ and $\pi_{{\rm E},\perp}$ are the components
parallel and perpendicular to the Earth acceleration at $t_{0, par}$ projected 
onto the plane of the sky. Equatorial north and east directions are indicated on the 
last panel.
{\bf Panel a)}
projection of all MCMC links for both $u_0>0$ and $u_0<0$ solutions. 
Parameter combinations that would yield unbound orbits (projected kinetic energy higher 
than the projected potential energy, Sec.~\ref{sec:orbit}) or would require negative 
blended light (Sec.~\ref{sec:posflux}) are excluded.
{\bf Panel b)}
the MCMC links are weighted by the lensing probability and the jacobian of 
the change from microlensing
parametrization to the physical space of distance, mass and 
projected velocities (Sec.~\ref{sec:gal}).
The result show the posterior one would derive assuming flat priors on 
$D_{\rm L}$, $\log M_{\rm host}$ and $\boldmath{v}_{{\rm L}, \perp}$.
{\bf Panel c)}
Same as the previous panel, but with an additional constraint from the mass-luminosity 
relation for the lens: the distance and the mass of the lens cannot yield
observed {\it I}-band magnitude that is brighter than the allowed amount of the 
blended light (Sec.~\ref{sec:ib}).
{\bf Panel d)}
all previous constraints together with the priors from the Galactic model, i.e.
density profiles of the disk and the bulge, velocities and velocity dispersions
of both populations and the mass function (see Sec.~\ref{sec:gal} for more details).
It is clear that there is no strong tension between the priors and the 
microlensing light curve solution.  
}
\end{figure*}

To illustrate this point, the left panel of Fig.~\ref{fig:prior} shows the values of $\pivec_\e$
that are allowed by the light curve model.
The remaining parameter space, is thus, rejected by the light curve.


\notetoeditor{(fig:prior) black and white for print, in color in the electronic version}

\subsection{The orbital motion of the lens}
\label{sec:orbit}
As mentioned earlier, we expect the planet to move in the orbit around its host,
and we take this into account by allowing the 
$\gamma_\parallel \equiv ds/dt/s_0$ 
and $\gamma_\perp \equiv d\alpha/dt$ to be non-zero.
In our modelling, while the best values of $\gamma_\perp$ stays around zero for the best models,
the $\gamma_\parallel$ moves to the negative side with the best fit value 
around -1.5 ${\rm yr}^{-1}$.
While this would indicate a fast orbital motion and near edge-on orbit, 
the $\chi^2$ improvement is only on the order of a few, and zero motion in this 
direction is still a viable solution, able to explain the shape of the light curve well.

Since, it is possible to evaluate mass and the projected separation
of the lensing system (see Section~\ref{sec:eqs}), following
\citet{dong07}, for each trial
set of parameters, we calculate the ratio of transverse kinetic to potential 
energy ($\beta = |{\rm KE}_\perp/{\rm PE}_\perp|$)  
and discard all that show $\beta>1$ as obviously
unphysical (or unbound) solutions. This reduces the allowed space of 
the orbital motion parameters to about $\gamma_\parallel = -0.4 \pm 0.4\, {\rm yr}^{-1}$
and $\gamma_\perp = 0.0 \pm 0.4\, {\rm rad}\,{\rm yr}^{-1}$.

For the set of physically bound solutions, the best $\chi^2$ 
improvement in respect to the static binary solution is only about $1.5$. 
Hence, the existence of the orbital motion 
in the system is not detected. At this point, one would argue 
that it is unnecessary to introduce those additional parameters to the
model, but rather, the final planetary system parameters should be
based on the static binary solution.
This might be, however, not always true.

Although, for the statistical model, adding additional degrees of freedom
and not getting enough improvement in the goodness of fit is 
considered counter-productive, it is only the case 
if the model is significantly disjoint from the physics. 
Here, we {\it expect} some amount of orbital motion in the system.

If we would look at the Jupiter's orbit face-on, the Jupiter would move
at a pace of $\sim 0.5\, {\rm rad}\, {\rm yr}^{-1}$. When it would be seen edge-on,
with the projected separation of 3 AU from the Sun, it would be getting closer
in the rate of $2.25\, {\rm AU}\, {\rm yr}^{-1}$, which corresponds to $\gamma_\parallel \sim 0.75\, {\rm yr}^{-1}$.

It is true, that the introduction of the 
additional parameters to 
the description of the event increases the uncertainties in the standard 
parameters, and thus, limit the predictive strength of the model.
However, this is actually a very welcome effect, as it reflects our ignorance
about the system. Avoiding such increase of the uncertainties 
might lead to 
the overestimated confidence in the results.

Here, for the inference of the planetary system physical parameters
we use the full set of fit variables and do not assume static binary
system.


\subsection{Resulting Microlensing Parameters}
Table~\ref{table:ulens} shows our best estimates of the event's microlensing parameters and 
the uncertainties derived from the MCMC sampling around the two 
solutions ($u_0>0$ and $u_0<0$). The model consists of the rotating binary 
lens, extended source star, and the relative lens-source motion affected by the annual parallax modulation.

Table~\ref{table:compar} provides some insights onto how the 
inclusion of parallax and orbital motion into the fits influences
the results. 

It is evident that five of the basic binary-lens parameters ($t_0$, $t_{\rm eff}$, $t_\e$, $s_0$, $\alpha_0$) 
have their uncertainties increased 3--4 times by allowing the parallax to vary within the microlensing fit.
The mass ratio ($q$) and the source-radius crossing time ($t_*$) are nearly unaffected.
While most of the affected parameters are in essence a nuisance parameters, the Einstein ring crossing time ($t_\e$)
enters proportionally into the $\theta_\e$ estimation (Eq.~\ref{eq:theta_rho}).
Note, that the uncertainty of the relative lens-source proper motion value is not impacted, since $\mu_{\rm geo}=\theta_*/t_*$.

The introduction of the orbital motion into the fit does not have a strong effects on the microlensing parameters in 
the particular case of the event discussed 
(with the exception of the parameters immediately related to the angle of the binary axis and the host-planet separation).
However, it is the case, that for certain events, the orbital motion and the parallax effects are highly correlated
\citep[eg.][Fig.~5]{skowron11}, and thus, the orbital motion might influence the results. In that sense, it 
is profitable to test ones models against such possibility.

Note also, that there is much less unmagnified flux (the \emph{blend} flux) allowed by the static binary model 
without the parallax than in all other models. This has implications onto the upper limit of the host star mass 
(as introduced in Sec.~\ref{sec:ib}) -- by relying only on the predictions of the simple static model 
one would reject solutions with higher mass (brighter) host stars that are perfectly allowed by the more complex models.

\begin{deluxetable}{l|rr|rr|rr|rr|}
\tablecaption{Lensing Parameters\label{table:ulens}}
\tablewidth{0pt}  
\tablecolumns{9}
\tabletypesize{\footnotesize}
\tablehead{
&  \multicolumn{4}{c|}{$u_0>0$ solution} &  \multicolumn{4}{c|}{$u_0<0$ solution}\\
\multicolumn{1}{l|}{parameter} &  \multicolumn{2}{c|}{raw MCMC} &  \multicolumn{2}{c|}{weighted} &  \multicolumn{2}{c|}{raw MCMC} &  \multicolumn{2}{c|}{weighted}
}
\startdata
\multicolumn{1}{l|}{$\chi^2/dof$} &  \multicolumn{2}{c|}{2595.43/2445} & \multicolumn{2}{c|}{\nodata} & \multicolumn{2}{c|}{2595.43/2445} & \multicolumn{2}{c|}{\nodata} \\
               $t_0$ (${\rm HJD}'$)   & $5673.81 $&$ \pm    0.18   $    & $5673.788$&$   \pm    0.084  $  &$5673.74 $&$\pm0.33           $ &$5673.687   $&$\pm    0.099  $\\
       $t_{\rm eff}$ (days)           & $  30.76 $&$_{-0.82}^{+0.61}$   & $  30.86 $&$_{-0.37}^{+0.27}$   &$ -30.9  $&$_{-1.4  }^{+1.1  }$ &$ -31.21    $&$_{-0.33}^{+0.39 }$\\
         $t_{\rm E}$ (days)           & $  34.20 $&$ \pm    0.74   $    & $  34.26 $&$   \pm    0.67   $  &$  34.2  $&$\pm1.3  $           &$  34.45    $&$\pm    0.55   $\\
               $t_*$ (days)           & $   0.594$&$ \pm    0.072  $    & $   0.599$&$   \pm    0.071  $  &$   0.602$&$\pm0.063$           &$   0.613   $&$\pm    0.068  $\\
                 $q$ ($\times 10^{-3}$)  & $   0.127$&$ \pm    0.025  $    & $   0.125$&$   \pm    0.023  $  &$   0.122$&$\pm0.025$           &$   0.118   $&$\pm    0.022  $\\
               $s_0$                  & $   1.691$&$ \pm    0.036  $    & $   1.694$&$   \pm    0.032  $  &$   1.699$&$\pm0.029$           &$   1.701   $&$\pm    0.028  $\\
          $\alpha_0$ ($\deg$)         & $-125.6  $&$ \pm    1.3    $    & $  -125.5$&$   \pm    1.2    $  &$ 125.3  $&$\pm1.4  $           &$ 125.3     $&$\pm    1.2    $\\
   $\pi_{{\rm E},N}$                  & $   0.07 $&$_{-0.12}^{+0.16}$   & $   0.048$&$_{-0.030}^{+0.036}$ &$  -0.03 $&$_{-0.31 }^{+0.31 }$ &$   0.042   $&$_{-0.061}^{+0.034}$\\
   $\pi_{{\rm E},E}$                  & $  -0.002$&$ \pm    0.033  $    & $  -0.009$&$   \pm    0.020  $  &$  -0.018$&$\pm0.029$           &$  -0.018   $&$\pm    0.017  $\\
        $d\alpha/dt$ (${\rm yr}^{-1}$)& $  -0.09 $&$ \pm    0.44   $    & $  -0.09 $&$   \pm    0.44   $  &$   0.07 $&$\pm0.41 $           &$   0.12    $&$\pm    0.43   $\\
             $ds/dt$ (${\rm yr}^{-1}$)& $  -0.56 $&$_{-0.52}^{+0.61}$   & $  -0.58 $&$_{-0.51}^{+0.61 }$  &$  -0.58 $&$_{-0.50 }^{+0.61 }$ &$  -0.62    $&$_{-0.51}^{+0.63 }$ \\ 
               $F_{\rm blend}$ & $0.78$&$_{-0.54}^{+0.90}$ & $0.79$&$_{-0.49}^{+0.70}$ & $0.64$&$_{-0.44}^{+0.70}$ & $0.68$&$_{-0.43}^{+0.62}$ \\ \hline
               $F_{\rm S} $    & $  11.53   $&$_{   -0.90   }^{   +0.54   }$ & $  11.51   $&$_{   -0.71   }^{   +0.49   }$ & $  11.67   $&$_{   -0.70   }^{   +0.44   }$ & $  11.62   $&$_{   -0.61   }^{   +0.43   }$ \\
               $u_0$                  & $   0.906$&$_{-0.037}^{+0.021}$ & $   0.905$&$_{-0.029}^{+0.019}$ &$  -0.911$&$_{-0.017}^{+0.028}$ &$  -0.909   $&$_{-0.017}^{+0.025}$ \\
               $\rho_*$ ($\times 10^{-3}$) & $17.4$&$\pm2.2$ & $17.5$&$\pm 2.1$ &$17.6$&$\pm 1.9$ &$17.8$&$\pm 2.0$ \\
               $\pi_{{\rm E}}$         & $0.113$&$_{-0.076}^{+0.139}$ & $0.055$&$_{-0.017}^{+0.034}$ & $0.21$&$_{-0.16}^{+0.22}$ & $0.055$&$_{-0.017}^{+0.032}$ 
\enddata
\tablecomments{
The parameters of the microlensing model -- both solutions with positive ($u_0>0$) and negative ($u_0<0$) impact parameters are shown.
  Raw results from the MCMC modeling are presented with the two constrains on the chain: blended light (``third light'') ought to be not negative (Sec.~\ref{sec:posflux}), and kinetic energy projected onto the plane of the sky has to be smaller than the projected potential energy (Sec.~\ref{sec:orbit}).
The ``weighted'' parameters are after the inclusion of the geometric weighting, 
all priors from the Galactic model (see Sec.~\ref{sec:gal}), 
the source proper motion measurement (Sec.~\ref{sec:pm}), 
and the limit on the lens brightness (not to exceed the observed blended flux, Sec.~\ref{sec:ib}). 
After weighting, the $u_0>0$ solution holds 68\% of weight, while $u_0<0$ solution only 32\%.
${\rm HJD}'={\rm HJD}-2450000$. $\alpha_0$ and $s_0$ denote projected binary axis angle and separation
for the epoch $t_{0,{\rm orb}}=5678.0$, respectively.
The reference position
for the definition of $t_0$ and $u_0$ is set as the center of mass of the lens system.
$u_0 = t_{\rm eff}/t_{\rm E}$,  $\rho_* = t_*/t_{\rm E}$ and $I_{\rm b}=18-2.5\log{F_{\rm b}}$.
Geocentric reference frame is set in respect to the Earth velocity at $t_{0,{\rm par}}=5678.0$.
}
\end{deluxetable}

\begin{deluxetable}{l|rl|rl|rl|rl|}
\tablecaption{Comparison of the Uncertainties in the Lensing Parameters for Various Models\label{table:compar}}
\tablewidth{0pt}  
\tablecolumns{9}
\tabletypesize{\footnotesize}
\tablehead{
\multicolumn{1}{l|}{parameter} &  \multicolumn{2}{c|}{static binary} &  \multicolumn{2}{c|}{orbiting binary} &  \multicolumn{2}{c|}{parallax-only} &  \multicolumn{2}{c|}{orbiting with parallax}
}
\startdata
\multicolumn{1}{l|}{$\chi^2$} &  \multicolumn{2}{c|}{2597.2} & \multicolumn{2}{c|}{2595.3} & \multicolumn{2}{c|}{2597.2} & \multicolumn{2}{c|}{2595.3} \\
$t_0$ (${\rm HJD}'$)          &  $5673.707$ & $   \pm    0.049      $ & $5673.694 $ & $  \pm    0.053        $ & $5673.82 $ & $\mathbf{\pm 0.17}$ (p)  & $5673.81  $ & $ \mathbf{\pm 0.18}$ (p)  \\
$t_{\rm eff}$ (days)          &  $  31.22 $ & $   \pm    0.16       $ & $  31.06  $ & $  \pm    0.25         $ & $  30.77 $ & $\mathbf{\pm 0.70}$ (p)  & $  30.70  $ & $ \mathbf{\pm 0.76}$ (p)  \\
$t_{\rm E}$ (days)            &  $  33.91 $ & $   \pm    0.26       $ & $  34.19  $ & $  \pm    0.45         $ & $  34.07 $ & $\mathbf{\pm 0.68}$ (p)  & $  34.20  $ & $ \mathbf{\pm 0.74}$  (p)  \\
$t_*$ (days)                  &  $   0.585$ & $   \pm    0.046      $ & $   0.592 $ & $  \pm    0.068        $ & $   0.576$ & $  \pm    0.054         $& $   0.594 $ & $ \pm    0.072          $ \\
$q$ ($\times 10^{-3}$)        &  $   0.116$ & $   \pm    0.016      $ & $   0.107 $ & $  \pm    0.017        $ & $   0.124$ & $  \pm    0.022         $& $   0.127 $ & $ \pm    0.025          $ \\
$s_0$                         &  $ 1.6871 $ & $ _{-0.0102}^{+0.0072}$ & $   1.705 $ & $\mathbf{_{-0.045}^{+0.025}}$(o)& $   1.676$ & $\mathbf{_{-0.025}^{+0.015}}$(p)& $   1.693 $ & $\mathbf{_{-0.037}^{+0.035}}$(po) \\
$\alpha_0$ ($\deg$)           &  $-124.90 $ & $   \pm    0.16       $ & $ -124.22  $ & $\mathbf{_{-1.54}^{+0.64}}$(o)& $-125.23 $ & $\mathbf{\pm 0.56}$ (p)& $-125.6   $ &$\mathbf{_{-1.3}^{+1.4}}$(po) \\
$\pi_{{\rm E},N}$             &     \multicolumn{2}{c|}{--}         &           \multicolumn{2}{c|}{--}      & $   0.08   $ & $  \pm    0.15          $& $   0.08  $ & $ \pm    0.15           $ \\
$\pi_{{\rm E},E}$             &     \multicolumn{2}{c|}{--}         &           \multicolumn{2}{c|}{--}      & $  -0.001$ & $  \pm    0.032           $& $  -0.002 $ & $ \pm    0.033          $ \\
$d\alpha/dt$ (${\rm yr}^{-1}$)&     \multicolumn{2}{c|}{--}         & $   0.31  $ & $_{-0.53}^{+0.22}      $ &           \multicolumn{2}{c|}{--}       & $  -0.09  $ & $\pm    0.44            $ \\
$ds/dt$ (${\rm yr}^{-1}$)     &     \multicolumn{2}{c|}{--}         & $  -0.49  $ & $_{-0.56}^{+0.79}      $ &           \multicolumn{2}{c|}{--}       & $  -0.52  $ & $\pm    0.56            $ \\
$F_{\rm blend}$ (up. limit)              &   \multicolumn{2}{c|}{0.89}         &   \multicolumn{2}{c|}{1.57}            &    \multicolumn{2}{c|}{2.11}          &       \multicolumn{2}{c|}{2.43}          
\enddata
\tablecomments{
The comparison of the uncertainties in the microlensing parameters for 4 models: lens as a static binary, orbiting binary, static binary with microlens parallax, and orbiting binary with parallax. 
The inclusion of the orbital motion has a significant effect on the binary separation ($s_0$) and the angle of the binary axis ($\alpha_0$) at the reference time $t_{0,{\rm orb}}=5678.0$. These values are marked in the table with: (o).
The introduction of the microlens parallax effect into the fit, while does not influence the best $\chi^2$ value, has a significant impact on the uncertainties in most of the parameters (marked with (p)) -- with an exception of source-radius crossing time ($t_*$) and mass ration ($q$).
Results show, that ignoring the parallax effect in the microlensing fit (even without significant gains in the goodness-of-fit) leads to an over-confidence in the estimation of the microlensing parameters. Ignoring the orbital motion of the lens, in this particular event, has smaller effect -- although in events where parallax is strongly correlated with the orbital motion, might influence the end results.
Last row shows 95\% upper limits on flux of the blend -- simple static binary model allows for much less blended light than more complex models, thus, might exclude more massive (brighter) lenses than are allowed by the other models.
(Only the parameters for solutions with positive impact parameter ($u_0>0$) are shown -- the other solution yield analogous results. Parameter definitions are as in Table~\ref{table:ulens}.)
}
\end{deluxetable}


\section{Constraining Physical Parameters of the System}
\label{sec:phys}

\subsection{Source star angular radius}
  
\label{sec:thetastar}
From the microlensing fit to the OGLE {\it I}-band and {\it V}-band data, we find 
the observed flux that was magnified in both bands. From that we find the
color of the source star to be: $((V-I)_S, I_S) = (1.829, 15.30)$.
With the method described by \cite{nataf10} we find the centroid of the
red clump stars in the 2 arc minute radius around the position of the event:
$((V-I)_{\rm RC}, I_{\rm RC}) = (1.850, 15.379)$.
The source star is 0.02 mag bluer and  0.08 mag fainter then the red clump,
therefore, it is most likely a K-type red clump star located in the Galactic bulge. 
The intrinsic color of the red clump centroid is $(V-I)_{{\rm RC},0} = 1.06$ \citep{bensby11}.
This yields the dereddened color of the source star:
\begin{equation}
 (V-I)_{S,0} = 1.06 - 0.02 = 1.04
\end{equation}
and the total reddening toward the Galactic bulge in this small region around the star:
\begin{equation}
  E_{(V-I)} = (V-I)_{\rm RC} - (V-I)_{{\rm RC},0} = 1.85 - 1.06 = 0.79.
\end{equation}
\citet{nataf13} provides the mean distance modulus of the
red clump stars toward our line of sight at ${\rm DM} = 14.52 \pm 0.23$
and unreddened absolute brightness of the red clump stars: $M_{I,RC,0} = -0.11$.
This allows us to calculate amount of extinction in the {\it I}-band towards the field:
\begin{equation}
  \label{eq:ai}
  A_I = I_{\rm RC} - {\rm DM} - M_{I,{\rm RC},0} = 15.38 - 14.521 +0.11 = 0.97,
\end{equation}
as well as in the {\it V}-band: $A_V = A_I + E_{(V-I)} = 1.76$.
Therefore, the extinction-free brightness of the source star is: 
\begin{gather}
  V_{0,S} = V_{S} - A_V = 17.13 - 1.76 = 15.37, \\
  I_{0,S} = I_{S} - A_I = 15.30 - 0.97 = 14.33.
\end{gather}
Following the discussion of the measurement uncertainties 
of the unreddened color and the extinction-free brightness 
of the microlensed sources in \citet[see Section 4.2]{skowron15},
we choose the color uncertainty to be 0.06 and 
brightness uncertainty to be 0.1 mag.

With the color-color relation from \citet{bessell88} 
and measured $(V-I)_{S,0}$ of the source star we find 
$(V-K)_{S,0} = 2.41 \pm 0.15$.
Then, the surface brightness calibration provided by \cite{kervella04b}, for the given $V_{S,0}$ and $(V-K)_{S,0}$
yields:
\begin{equation}
  \theta_{*} = 6.08 \pm 0.65\, \mu\mathrm{as}.
\end{equation}

With use of the Eq.~\ref{eq:theta_rho}, 
this gives the angular Einstein ring radius of:
\begin{equation}
  \theta_\e = 0.337 \pm 0.053\, \mas.
\end{equation}

\subsection{The measurement of the source proper motion}
\label{sec:pm}
We note that, since not-magnified light (blending) in this event is
insignificant when compared to the light from the source star 
($F_{blend}/F_S = 0.07 \pm 0.07$), and
since this region of the sky  was monitored by the OGLE-III for eight
years, it is possible to measure the proper motion of the source
star and, thus, further constrain the microlensing solution.

Analogously to the process described by \citet{poleski12}, 
we measure the positions
of all stars in the $7' \times 7'$ region around the event
on all science frames of the field \emph{BLG196.5} of the OGLE-III survey,
and derive their proper motions.
These motions are relative to some mean frame of motion based on brighter 
stars in the field and do not have physical meaning. Under the assumption
that the mean proper motion of the Galactic bulge stars toward this direction
($(l, b) = (1.7,-3.5)$) is close to being stationary in respect to the
Galaxy, we calibrate motion of all stars in the field to the sample of
known Galactic bulge stars.
On the Color-Magnitude Diagram we choose 1441 stars in the narrow
ellipse around the centroid of the red clump stars region and analyze their motion.
Typical uncertainties of the individual proper motions in the sample are $0.4\,\masyr$.
The measured mean motion of the red clump giant sample is $(-1.00, 0.23) \pm (2.78, 2.91)\, \masyr$
in the $(l, b)$ direction. We see, that
our fiducial reference frame is moving mainly in the direction of Galactic rotation.
This is actually expected, since this frame is attached to the mix of bright bulge and disk stars
in the field. We also measure the bugle velocity dispersion in this direction to be
$\sim 2.8\,\masyr$ which translates to $\sim 106\,\kms$ -- in 
agreement with the theoretical models.

The proper motion of the source star in the event MOA 2011-BLG-028 is measured
to be:
\begin{equation}
  \bmu_{0,S} = (-2.79, -1.96) - (-1.00, 0.23) = (-1.79, -2.19) \pm 0.37\; \masyr
\end{equation}
in the $(l, b)$ direction, and is relative to the mean motion of the Galactic bulge stars
(which we indicate with an index ``0'').

\subsection{Upper limit on the lens brightness}
\label{sec:ib}
Each trial solution in the MCMC process consists of 
a set of microlensing parameters (describing the 
magnification changes) and a blend flux ($F_{\rm blend}$).
The remaining light from the OGLE-IV baseline is 
the source star flux ($F_{\rm S} = F_0 - F_{\rm blend}$) and is needed in order to compare theoretical
magnification with the observed flux. 
As we discussed in Sec.~\ref{sec:posflux} we 
require that $F_{\rm blend} \ge 0$ by rejecting trials 
with negative blend flux.

In the case were the angular Einstein ring radius 
is measured (Sec.~\ref{sec:thetastar}), the solutions
with a very low value of $\pi_\e$, 
points to a very high mass of the lens
(Eq.~\ref{eq:mass}). In the case of a main-sequence 
star, this would lead to a very bright lens.
There is a limit, in which, the brightness of such lens would
definitely overcome the blended light seen in the light 
curve.

For each link in the MCMC chain, we calculate the distance to the lens (Eq.~\ref{eq:pirel}) 
and from it, its distance modulus
(${\rm DM}_{\rm L}$), as well as the mass of the star 
($M_{\rm host}$).
We use the main-sequence isochrones calibrated 
by \cite{an07} and estimate the absolute {\it I}-band 
magnitude ($M_{I,{\rm iso}}$) for the given stellar
mass.
The lower limit on the brightness of the
host star ($I_{\rm limit}$) is set under the 
assumption that the lens is located behind 
all the dust, which we measure in front of the
source star, and is equal to:
\begin{equation}
  I_{\rm limit} = M_{I,{\rm iso}} + {\rm DM}_{\rm L} + A_I
\end{equation}
where $A_I$ is taken from the Eq.~(\ref{eq:ai}).
After the conversion to fluxes ($I = 18 - 2.5 \log F$),
we require that:
\begin{equation}
  F_{\rm limit} \le F_{\rm blend},
\end{equation}
not to allow host star to be brighter than the amount 
of blended light seen in the light curve.

Uncertainty of the mass estimation for each MCMC link 
reflects the uncertainty in the $\theta_\e$. In order to
take this into account, we convolve the 
less brightness limit together with the $\theta_\e$ uncertainty.
This process produces a numeric weight, with which, the particular
MCMC link enters into the final considerations.

\subsection{Priors from the Galactic model}
\label{sec:gal}
The measurement of the angular
Einstein radius provides one dimensional relation
between the lens mass and its distance. This degeneracy
can be broken by the measurement of the microlens parallax vector
length. Unfortunately in the case discussed here, we are unable
to detect clear influence of the Earth motion
onto the light curve. In other words, the best solution is close to the point $\pivec_\e = 0$
(see Table~\ref{table:ulens} and the first panel of Fig.~\ref{fig:prior}).
We opt to estimate the physical parameters of the system
by employing priors from the Galactic model and, as discussed above,
additional information and limits.

We use Jacobian from \citet[Eq.~(18)]{batista11} to move the 
parametrization from the microlensing variables to the 
physical parameters -- the ones we have priors expressed in.
This formula also corrects for the geometric effects
of viewing angle, and introduces weighting for the lensing
probability (which is proportional to the size of the 
Einstein ring and lens velocity).

For the mass density model of the Galactic bulge
we employ E3 model by \citet{cao13} fitted 
in the region $|l|<4^\circ$, $|b|<4^\circ$ to the 
red clump giants count, mean distance
moduli and dispersions measured by \cite{nataf13}
from the OGLE-III data. 
We use Galactic disk mass density model from \citet{hangould03}
that is based on \citet{zheng01} $sech^2$ model.
We also use mass functions described in the Appendix B2 of 
\citet{dominik06}.

The velocity of the Galactic bulge stars is assumed to be
on average zero $(v_l, v_b) = (0, 0)\,\kms$ (with respect to the Galaxy)
with the dispersion of 100 $\kms$ in
both directions (i.e., direction of the Galactic rotation, and 
toward the Galactic north). Disk rotation velocity is taken as 220 $\kms$
with the dispersions of 30 and 20 $\kms$ in $(l, b)$
directions, respectively. We also consider an asymmetric drift of 10 $\kms$
to account that the disk stars, on average, rotate a little 
slower that the Galactic disk gas.

Table~\ref{table:prior} presents the influence the different
weighing and limits have on the expected distance, mass and velocity
of the lensing system.
Figure~\ref{fig:prior} shows this in terms of microlens parallax vector.
We see, that the lens is most probably located in the Galactic
bulge and is a moderately-massive star. Also, there is no strong tension between 
the final parameters of the system and the raw light curve preference.

\begin{deluxetable}{l|rr|rr|rr|}
\tablecaption{Priors' influence on Mass and Distance\label{table:prior}}
\tablewidth{0pt}  
\tablecolumns{7}
\tabletypesize{\footnotesize}
\tablehead{\multicolumn{1}{l|}{} &  \multicolumn{2}{c|}{$M_{\rm host}$} &  \multicolumn{2}{c|}{$D_{\rm L}$} &  \multicolumn{2}{c|}{$v_{\perp, 0, {\rm L}, l}$}\\ \multicolumn{1}{l|}{choice of priors} &  \multicolumn{2}{c|}{$(M_{\odot})$} &  \multicolumn{2}{c|}{(kpc)} &  \multicolumn{2}{c|}{$(\mathrm{km}\,\mathrm{s}^{-1})$}}
\startdata
no priors (flat) and no cutoff for lens brightness                  & $1.39 $&$_{-0.75}^{+0.95}$ & $7.32 $&$_{-0.92}^{+0.80}$ & $-48 $&$_{-120}^{+130}$ \\
no priors (flat on distance, $\log M_{\rm host}$ and velocities)    & $0.78 $&$\pm 0.35        $ & $6.90 $&$_{-1.02}^{+0.85}$ & $7 $&$_{-155}^{ +93}$ \\
with prior on galactic density ($D_{\rm L}$)                        & $0.92 $&$\pm 0.31        $ & $7.42 $&$_{-0.60}^{+0.54}$ & $-28 $&$_{-140}^{+100}$ \\
with priors on galactic density and velocities in the Galaxy        & $0.93 $&$\pm 0.31        $ & $7.44 $&$\pm 0.65        $ & $30 $&$_{-120}^{ +48}$ \\
with priors on galactic density, $\log M_{\rm host}$ and velocities ($v_\perp$) & $0.75 $&$_{-0.30}^{+0.36}$ & $7.38 $&$_{-0.62}^{+0.53}$ & $40 $&$_{-119}^{ +44}$ \\
\hline
\enddata
\tablecomments{
The evolution of the chosen physical parameters of the system depending on the choice of priors. 
$M_{\rm host}$ is a mass of the host star, 
$D_{\rm L}$ is a distance to the lensing system 
and $v_{\perp, {\rm L}, l}$ is the component of the transverse velocity of 
the lens in the direction of Galactic rotation. 
Mass of the planet is $10^{-4}$ of its host's mass.
In our Galactic model, bulge stars on average move with $v_{\perp, l} \sim (0 \pm 100)\;{\rm km}\,{\rm s}^{-1}$ 
and disk stars move $\sim (210 \pm 30)\;{\rm km}\,{\rm s}^{-1}$. 
We assume priors on the distance based on the Galactic density models, 
priors on the lens transverse velocity based on assumed velocities and velocity 
dispersion of the disk and the bulge populations in the Galaxy, 
and priors on the lens mass using disk and bulge mass functions (see Sec.~\ref{sec:gal} for details). 
In all rows, except the first one, we assume that the lens cannot be brighter 
than the amount of the additional/unmagnified light (blend) seen during 
the microlensing event (Sec.~\ref{sec:ib}).
}
\end{deluxetable}

\section{Results and discussion}
\label{sec:results}
Based on the light curve analysis, source star proper motion 
and priors coming from the expected lens population, we 
are able to provide assessment of the physical parameters of the
planetary system. These are presented in Table~\ref{table:parms}.
The planet is a Neptune-class planet on the orbit of $\sim 0.75\,M_\odot$ Galactic bulge star
located approximately 7.4 kpc in the direction of $(l,b)=(1.7^\circ,-3.5^\circ)$.
At the time of the lensing event, the projected host-planet separation was
$4.2 \pm 0.6\, {\rm AU}$.


\begin{deluxetable}{lrl}
\tablecaption{Planetary System Parameters\label{table:parms}}
\tablewidth{0.5\textwidth}
\tablecolumns{3}
\tablehead{\colhead{quantity} & \multicolumn{2}{c}{final estimates}}
\startdata
$M_{\rm planet}$ ($M_\earth$)       & $  30       $&$_{  -12      }^{  +16      }$    \\
$M_{\rm host}$ ($M_\odot$)         & $   0.75    $&$_{   -0.30   }^{   +0.35   }$    \\
$D_{\rm L}$ (kpc)               & $   7.38    $&$_{   -0.62   }^{   +0.52   }$   \\
$a_\perp$ (AU)                  & $   4.14    $&$\pm    0.64   $  \\
$\theta_{\rm E}$ (mas)          & $   0.337   $&$\pm    0.053  $   \\
$\mu_{\rm geo}$ ($\masyr$)      & $   3.59    $&$\pm    0.58   $    \\
$\mu_{0,{\rm L}, l}$ ($\masyr$) & $   0.68    $&$_{   -3.24   }^{   +0.98   }$ \\
$\mu_{0,{\rm L}, b}$ ($\masyr$) & $  -0.1     $&$_{   -1.3    }^{   +1.0    }$ \\
$\mu_{0,{\rm S}, l}$ ($\masyr$) & $   -1.79   $&$\pm 0.37 $ \\
$\mu_{0,{\rm S}, b}$ ($\masyr$) & $   2.19    $&$\pm 0.37 $
\enddata
\tablecomments{
The physical parameters of the lensing system:
mass of the planet in Earth masses ($M_{\rm planet}$),
mass of the host star ($M_{\rm host}$),
distance to the lensing system ($D_{\rm L}$),
and projected star-planet separation ($a_\perp$),
angular Einstein radius ($\theta_{\rm E}$),
relative lens-source proper motion in the geocentric reference frame ($\mu_{\rm geo}$) for time $t_{0,par}$,
heliocentric lens ($\mu_{0, {\rm L}}$) and source ($\mu_{0, {\rm S}}$) 
proper motion in respect to the bulk motion of the Galactic bulge stars.
The presented values take into account volume effects, as well, as priors 
from the Galactic model: stellar density, velocities, and the mass function (Sec.~\ref{sec:gal}). 
Since the source proper motion was measured astrometrically (Sec.~\ref{sec:pm}), 
it is possible to estimate the lens proper motion.
We provide derived values together with standard deviations or 68\% confidence limits,
where appropriate.
}
\end{deluxetable}


The source star moves $2.8\, \masyr$ in respect to the mean proper 
motion of the bulge stars in the field (see Sec.~\ref{sec:pm}), 
and the relative lens-source proper motion measured from the 
light curve is $3.6\, \masyr$. Since the majority of the bulge stars have
proper motions within $1\, \masyr$ from the mean,  the
above values are highly compatible with the lens being a typical member 
of the bulge population. 
On the other hand, the source is moving
$\sim -1.8\, \masyr$ when projected onto the direction of disk rotation,
so for the typical disk lens, we would expect the relative proper motion 
of the microlensing event to be significantly higher ($\sim 7\, \masyr$) 
than the value actually measured from the light curve ($\theta_*/t_* = 3.6\, \masyr$).

The most likely location of the source star is slightly behind the mean 
distance to the bulge (which is 8 kpc in this field, \citealt{nataf13}) 
at $8.61 \pm 0.64 \kpc$. This 
is because the lens and the source star are drawn from the same bulge density
profile, and at the same time, the lens must be in front of the source.
Also, the probability of lensing by the star in short distance from the source 
is low.

\subsection{Follow-up observations}
The extinction in $K_s$-band in the discussed direction is estimated by \citet{gonzalez11}
to be $A_{Ks} = 0.16$. 
We take the main-sequence isochrones calibrated by \cite{an07} 
to find the absolute {\it K$_s$}-band
magnitude ($M_{K_s,{\rm iso}}$) for the given
mass of the host star. This, with the estimation of the distance to the lens,
gives the observed magnitude: $K_s = M_{K_s,{\rm iso}} + {\rm DM}_{\rm L} + A_{Ks}$.
This way we find, that the most likely value of the host star's observed brightness
is $K_s = 19.0 \pm 1.4$.

In April 2011 the lens and the source were in near-perfect alignment 
($\le 0.3$ mas).
In 10 years from now, the separation of the planetary system from the 
source giant will be 50 mas -- in theory, easy to separate
using the ground based AO system or the Space Telescope.  
Unfortunately, the observed brightness of the source star,
a red clump giant toward the Baade Window, is approximately $K_s = 13$
-- this is hundreds of times brighter than the expected 
brightness of the host star of the planetary system.
This makes any follow-up observations very challenging.

As an example, \citet{pietrukowicz12} 
did measured the displaced position of the lens with $K_s=20.6$
from the source star with $K_s=17.4$ at the separation of 125 mas
with 37 minutes integration at VLT NACO ($20 \times 110s$).
Therefore, the planetary system host (MOA 2011-BLG-028La),
if sufficiently separated from the source star, can be detected from the ground. 
The expected contrast of 6 mag between the lens and the source
is much bigger than 3.2 mag in the case studied by \citet{pietrukowicz12},
requiring longer waiting time before the follow-up observation could be performed,
or requiring the use of a space-based facility.

If, however, the host star is seen in the future, 
the measurement of its brightness could narrow the mass
and distance estimation for this planetary system \citep[eg.][]{bennett10,janczak10,fukui15}.
Also, the actual measurement of the lens-source separation,
and hence, the proper motion,
can serve as a useful cross-check for the evaluated here
angular size of the Einstein radius, since $\theta_\e = \mu t_\e$
\citep[see also Sec.~4.4 of][]{gaudi12}.


\section{Conclusions}
\label{sec:conclusions}

The microlensing event MOA 2011-BLG-028 is a low magnification 
that peaked in April 2011 in the direction of the Galactic bulge. 
The source star is a red clump giant, while the lens is most likely 
a main sequence star also in the Galactic bulge.

The dense observational coverage of the MOA 2011-BLG-028 microlensing event
allowed us to confirm existence of a planetary companion to the main lensing
body. The mass ratio is accurately measured to be $(1.2 \pm 0.2) \times 10^{-4}$
indicating a Neptune-class planet.

The finite source effects seen in the light curve allowed us to measure 
the relative lens-source proper motion. The light curve does not allow
for the microlens parallax measurement, however, some limits on this 
vector value exists.

The low amount of blended light in the light curve and the decade long OGLE
monitoring of the field allowed us to measure the source star proper motion.
We use these data as an additional argument for the location
of the planetary system inside the Galactic bulge.

We derive the expected physical parameters of the planetary system
with aid of the Galactic model density and the velocity distributions.
We also weigh our results with the mass function of the potential 
lenses and the expected lensing rate.

Low-mass, as well as, moderate-mass main-sequence stars are 
allowed by the fits. While the moderate-mass and high-mass lenses are preferred
by the Galactic density arguments, lenses with higher mass than
$\sim 1.3 M_\odot$ are rejected as the light from them would 
be clearly detected, and thus, are incompatible with observations.

We test our predictions based on the simple static binary microlensing 
model, as well as more complex models including the parallax and the orbital motion.
While none of these effects are proven to be detected in the light curve, 
a priori, we expect they could have influenced it,
as both effects surely are present in the physical reality of all Galactic microlensing events.
We see, that the fits not allowing for the parallax claim 3--4 times smaller
uncertainties of the basic microlensing parameters.
Also, the static binary model without the parallax more stringently 
rejects brighter lenses than more complex models. 
While the current mass and distance estimations of the planetary system
have very wide uncertainties, and it is not proven that the 
inclusion of the orbital motion is important in this event, 
the Bayesian analysis with the results of the no-parallax static binary model 
predicts the lensing system that is $\sim 30\%$ lighter and slightly 
closer than the system predicted using the models that include the 
parallax effect. 

The host is a $(0.8 \pm 0.3)\, M_\odot$ star located
$(7.3 \pm 0.7)\, \kpc$ away from the Sun in the direction of the Galactic bulge
and is hosting a 12-60 $M_\earth$ planet on $\sim 3-5$ AU orbit.

The considerable distance to the planetary system and 
projected proximity of the bright giant star makes
it a challenging target for the future follow-up observations.

\acknowledgments
\noindent{\bf Acknowledgments:} 
The OGLE project has received funding from the National Science Centre,
Poland, grant MAESTRO 2014/14/A/ST9/00121 to AU.
JS was partially supported by the Space Exploration Research Fund of The Ohio State University.
This research was partly supported by the Polish Ministry of Science
and Higher Education (MNiSW) through the program ``Iuventus Plus''
award No. IP2011 026771.
We thank Prof. M.~Kubiak for contributing observations to this work
as the OGLE Project member.
TS acknowledges the financial support from the JSPS, JSPS23103002,
JSPS24253004 and JSPS26247023. 
The MOA project is supported by the grant JSPS25103508 and 23340064.
Operation of the Danish 1.54m telescope at ESO’s La Silla observatory 
was supported by The Danish Council for Independent Research, 
Natural Sciences, and by Centre for Star and Planet Formation. 
The MiNDSTEp monitoring campaign is powered by 
ARTEMiS (Automated  Terrestrial Exoplanet Microlensing Search; \citealt{dominik08}). 
NJR is a Royal Society of New Zealand Rutherford Discovery Fellow.
JS wishes to explicitly thank Prof. Andy Gould for his valuable involvement into 
the initial phases of this project.

\end{document}